%% file: DUST_mnras_def.tex
\documentclass[useAMS,usenatbib]{mn2e}
\usepackage{graphicx}
\usepackage{txfonts}
\usepackage{url}
\usepackage[dvips,usenames]{color}
\input{journalDefs}

\input{emulateMNRAS}
\newcommand{\sub}[1]{\mbox{$_{\rm #1}$}}

\newcommand{\Teff}{\mbox{$T\sub{eff}$}}

\newcommand{\beq}{\begin{equation}}
\newcommand{\eeq}{\end{equation}}
\newcommand{\beqa}{\begin{eqnarray}}
\newcommand{\eeqa}{\end{eqnarray}}
\newcommand{\benu}{\begin{enumerate}}
\newcommand{\eenu}{\end{enumerate}}
\newcommand{\bite}{\begin{itemize}}
\newcommand{\eite}{\end{itemize}}
\newcommand{\bdes}{\begin{description}}
\newcommand{\edes}{\end{description}}

\newcommand{\comment}[1]{}

\title[Dust in TP-AGB Stars]{Evolution of Thermally Pulsing Asymptotic Giant Branch
Stars II. Dust production at varying metallicity}

\author[Nanni et al.]{Ambra Nanni$^1$, Alessandro Bressan$^1$, Paola Marigo$^2$,
  L\'eo Girardi$^{3}$
  \\
  $^1$ SISSA, via Bonomea 265, I-34136 Trieste, Italy \\
  $^2$ Dipartimento di Fisica e Astronomia Galileo Galilei,
  Universit\`a di Padova, Vicolo dell'Osservatorio 3, I-35122 Padova, Italy \\
  $^3$ Osservatorio Astronomico di Padova, Vicolo dell'Osservatorio 5,
  I-35122 Padova, Italy \\
}

\begin{document}

\date{Received 8 March 2013 / Accepted 25 June 2013}

\pagerange{\pageref{firstpage}--\pageref{lastpage}} \pubyear{2013}

\maketitle

\label{firstpage}

\begin{abstract}
We present the dust ejecta of the new stellar models  for the
Thermally Pulsing Asymptotic Giant Branch (TP-AGB) phase
computed with the \texttt{COLIBRI} code.
 We use a formalism of dust growth coupled with a stationary wind
for both M and C-stars.
In the original version of this formalism,
the most efficient destruction process of silicate dust in M-giants
is chemisputtering by H$_2$ molecules. For these stars
we find that dust grains can only form at relatively large radial distances ($r \sim 5\, R_*$),
where they cannot be efficiently accelerated, in agreement with other investigations.
In the light of recent laboratory results,
we also consider the alternative case that the condensation temperature
of silicates  is determined only by the competition between growth and free evaporation
processes (i.e. no chemisputtering).
With this latter approach we obtain dust condensation temperatures
that are significantly higher (up to $T_{\rm cond}\sim$1400~K) than those
found when chemisputtering is included ($T_{\rm cond}\sim$900~K), and in better
agreement with  condensation experiments.
As a consequence, silicate grains can remain stable in inner regions of the circumstellar envelopes
($r \sim 2\, R_*$), where they can rapidly grow and can be efficiently accelerated.
With this modification, our models nicely reproduce
the observed trend between terminal velocities
and mass loss rates of Galactic M-giants.

For C-stars the formalism is based on the homogeneous growth scheme where
the key role is played by the carbon over oxygen excess.
The models reproduce fairly well the terminal velocities of Galactic stars
and there is no need to invoke changes in the standard assumptions.
At decreasing metallicity the carbon excess becomes more pronounced and
the efficiency of dust formation increases. This trend could be in tension with
recent observational evidence in favour of a decreasing efficiency,
at decreasing metallicity. If confirmed by more observational data,
it would indicate that either the amount of the carbon excess, determined by
the complex interplay between mass loss, third~dredge-up and hot bottom burning, or
the homogeneous growth scheme should be revised.
Finally, we analyze the differences in the total dust production of M-stars that
arise from the use of the two approaches (i.e. with or without chemisputtering).
We find that, in spite of the differences in the expected dust stratification,
for a given set of TP-AGB models, the ejecta are only weakly sensitive
to the specific assumption.
This work also shows that the properties  of  TP-AGB circumstellar envelopes are
important diagnostic tools that may be profitably added to the traditional
calibrators for setting further constraints on this complex phase of stellar evolution.
\end{abstract}

\begin{keywords}
stars: AGB and post-AGB - stars: mass loss - stars: winds, outflows - circumstellar matter -
dust, extinction
\end{keywords}

\section{Introduction}
During the Asymptotic Giant Branch (AGB) phase, stars with initial masses
in the range $0.8\la M\la 6-8$~M$_{\odot}$ lose their
envelopes at typical rates
between 10$^{-8}$~M$_\odot$yr$^{-1}$ and few 10$^{-5}$~M$_\odot$yr$^{-1}$, polluting the
Interstellar Medium (ISM) with metals, partially condensed into dust.
Direct estimates of the amount of AGB dust and its mineralogy are provided by
mid and far-infrared observations,
both in our galaxy and in the nearby ones \citep{Knapp85, Matsuura09, Matsuura12}.

Comparing the dust mass loss rates derived from far-infrared observations with the
gas mass loss rates obtained from CO observations for Galactic AGB stars,
\citet{Knapp85} have found typical dust-to-gas ratios of $\sim$6~$\times$~10$^{-3}$
for oxygen-rich (M) stars, mainly in the form of
amorphous silicates,  and $\sim$10$^{-3}$ for carbon-rich (C) stars,
mainly as amorphous carbon.
In particular they have estimated that, at solar metallicity, a large fraction  of the silicon
should condense into dust in the circumstellar envelopes (CSEs) of these stars.
AGB stars were soon recognized as the main stellar dust producers,
compared to other sources like  Supernovae (SNe), Red Supergiant
and Wolf-Rayet stars \citep{Gehrz89}. This picture becomes more complex
considering that,
once dust is injected into the ISM, it is subject to many processes that
can alter significantly its abundance and composition \citep{Draine03}.

Moving beyond the Galactic context,  it is now well established that the stellar dust
has a far-reaching relevance,  being a critical element
in the interpretation of extraGalactic observations up to the far Universe.
The study of the Spectral Energy Distribution (SED) of galaxies
and quasars at high redshifts shows that even very young objects possess
large dust reservoirs \citep{Lilly99, Eales00, Bertoldi03, Robson04, Beelen06, Dwek11}.
On the theoretical side, many efforts have been made in order to  model dust evolution
in local galaxies \citep{Dwek98, Calura08, Zhukovska08, Piovan11a, Piovan11b, Boyer12, Boyer13},
and to explain the presence of large
amounts of dust at early epochs, when the dust and chemical enrichment time-scale was only a fraction of a Gyr
\citep{Dwek07, Valiante09,Mattsson10gal, Valiante11, Dwek11, Gall11, Pipino11a, Pipino11b, Michalowski11, Yamasawa11}.
Several authors suggest that at high redshifts the major dust contributors should
be SNe  because of their short lifetimes with respect to those of AGB stars
\citep{Maiolino04, Marchenko06}.  However, SNe play also an important role
in dust destruction since they produce shocks and high energy particles, so that
their  dust contribution is currently still controversial \citep{Todini01, Nozawa03, Sugerman06}.
On the other hand, recent studies have concluded
that the observed dust at these redshifts should be ascribed to the ejecta of
stars more massive than 3~M$_\odot$, with SNe confined to
a secondary contribution \citep{Dwek07, Valiante09, Dwek11}.
In this framework,
it is therefore crucial to know how stellar dust ejecta,
and in particular those of intermediate-mass stars,
depend on metallicity since, at least in the early phases of galaxy evolution,
the environment conditions  were likely quite different from those
in the present local Universe.

Another key motivation behind this study is the need for calibrating uncertain parameters that are commonly used
in stellar evolution calculations, such as  the efficiency of mass loss.
New sets of Thermally Pulsing Asymptotic Giant Branch (TP-AGB)
evolutionary tracks, for a wide range of initial masses and metallicities,
have been recently presented
by our group \citep{marigoetal13}.
It is now necessary to extend the study to those AGB stellar observables,
e.g. their mid-infrared colours,
that critically depend on the mass loss rates.
Stellar isochrones in mid-IR bands have already been calculated
\citep{Bressan98, Marigo_etal08},
but these studies do not rely on direct calculations of the properties of dust in the CSEs.
With this work we aim at predicting all the basic quantities needed
for stellar population studies in the mid-infrared,
i.e. dust ejecta, dust-to-gas ratio, composition, and outflow velocity,
as a function of stellar parameters.

Detailed modeling of dust formation within CSEs of AGB stars has already
been carried out by several authors.
On one side,  full hydrodynamical computations of dust formation,
coupling radiative transfer with pulsations and induced-shocks,
describe the complex interaction between radiation and dust grains
\citep{Bowen91, Fleischer91, Lodders99, Cherchneff00, Winters00, Elitzur01, Jeong03, Hoefner03, Hofner08}.
A different approach is that of describing the
dust growth in an expanding envelope,
under the stationary-wind approximation as in \citet{FG06} (hereafter FG06) and in \citet{GS99}.
Unlike the former approach, the latter is not fully consistent since,
for example, it cannot predict the mass loss rate as a function of stellar parameters.
However, at present this simplified method is the only feasible way to couple
dust formation with stellar evolution calculations that, especially for the
TP-AGB phase, involve a large number models.
Furthermore, this method can be useful
for testing the effects of some critical assumptions.
As a matter of fact, there are still several uncertainties that affect the theory of dust formation.
In particular the dust condensation temperature of silicates obtained from
the original and widely used models by \citet{GS99} and  FG06,  $T_{\rm cond}\leq$1000~K,
is significantly lower than  that measured in more recent laboratory experiments, $T_{\rm cond}\sim$1350~K \citep{Nagahara09}
and this may affect significantly the results of the dust formation models.

Since the new TP-AGB tracks by \citet{marigoetal13} span a wide range of masses and metallicities and,
following in a very detailed way the various physical processes inside the star,
comprehend thousands of models, we opt for
the more agile approach of FG06.
Thanks to the flexibility of this method,
we will also investigate the effects of using a new
criterion for the determination of the condensation temperature of silicates.
In particular, we will compare our predictions  with the observed velocities
of Galactic AGB stars, and analyse the impact
of the different assumptions on the resulting dust ejecta.

The paper is organized as follows.
In Section~\ref{tpagbmodels} we summarize the main characteristics
of the underlying TP-AGB tracks.
The basic equations of the wind model are presented
in Section~\ref{sec_wind}.
In Section~\ref{sec_growth} we discuss
the equations governing the dust growth,
with particular attention to the dust destruction processes.
We illustrate two different
methods to compute the condensation temperature of silicates.
In Section~\ref{sec_res} we apply both methods to the TP-AGB models,
 and compare the results with observed terminal velocities.
Dust ejecta of different masses and metallicities are provided in
Section~\ref{sec_dust_ejecta}.
Finally, the results are discussed in Section~\ref{sec_discussion}.

\section{The TP-AGB models}\label{tpagbmodels}
The TP-AGB evolutionary models are computed with the new code \texttt{
COLIBRI}, described in \citet{marigoetal13} to which the reader should
refer for all details. In brief, for each combination of initial
stellar mass ($M$) and initial metallicity ($Z$), the characteristic quantities at the
first thermal pulse, obtained from the \texttt{PARSEC} database
of
full stellar models \citep{Bressanetal12}, are fed as initial
conditions into \texttt{COLIBRI}, that calculates the whole TP-AGB
evolution until the entire envelope is lost by stellar
winds.

It is worth emphasizing that, compared to purely synthetic TP-AGB
codes, \texttt{COLIBRI} relaxes a significant part of their previous analytic
formalism in favour of  detailed physics applied to a complete
envelope model, in which the stellar structure equations are integrated
from the atmosphere down to the bottom of the hydrogen-burning shell.
As a consequence, both the hot-bottom burning (HBB) energetics and
nucleosynthesis, as well as the basic changes in envelope structure --
including effective temperature and radius -- can be followed with the
same richness of detail as in full models, and even more accurately.

In fact, a unique feature of \texttt{COLIBRI}, which is of particular
importance for the present work, is
the first {\em on-the-fly} computation ever with the \texttt{\AE SOPUS} code
\citep{MarigoAringer_09} of i) the chemistry for roughly $300$ atoms  and
$500$ molecular species and ii) gas opacities throughout the atmosphere
and the deep envelope at each time step during the TP-AGB phase.
This new technique assures a consistent coupling of the envelope
structure with its chemical composition, that may significantly
change due to the third dredge-up and HBB processes.

In particular, with \texttt{COLIBRI} we are able to follow in detail
the evolution of the surface C/O ratio, which is known to produce  a
dramatic impact on molecular chemistry, opacity, and
effective temperature every time it crosses the critical region around unity
\citep{Marigo_02, MarigoAringer_09}. In turn, the C/O ratio plays a paramount
role in determining the chemical and physical
properties of the dust, as we discuss in this work.
We just mention here that in M-stars (C/O$<1$) the main dust species are
amorphous silicates, quartz (SiO$_2$) and
corundum (Al$_2$O$_3$) \citep{Tielens98, Ossenkopf92}.
On the other hand, in C-stars (C/O$>1$)
the dust produced is
predominantly amorphous carbon and silicon carbide (SiC) \citep{Groenewegen98}.

\begin{figure*}
\centering
\includegraphics[angle=90,width=0.95\textwidth]{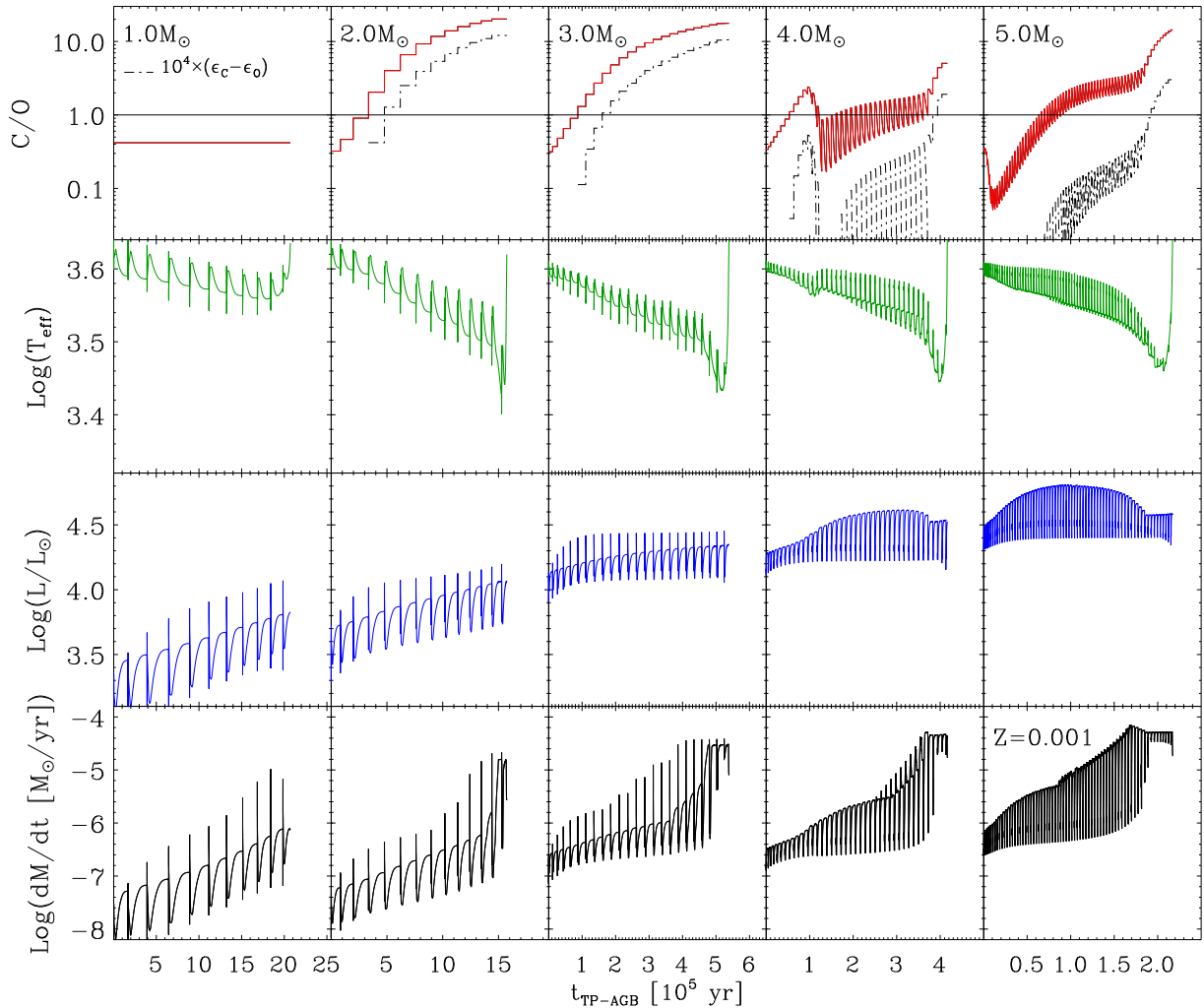}
\caption{Evolution of  surface C/O, carbon excess $\epsilon_{\rm C}$-$\epsilon_{\rm O}$ (only when positive),
 effective temperature, luminosity, and
mass loss rate during the whole TP-AGB phase
of a few selected models with initial metallicity $Z=0.001$, computed
with the \texttt{COLIBRI} code \citep{marigoetal13}.
These quantities are the key
input stellar parameters for our dust growth model. Time is counted from the first thermal
pulse. Note that effective temperature and luminosity are obtained from the solution of the full set
of the stellar structure equations, and not from fitting relations as usually
done in synthetic TP-AGB models. The trends of the C/O ratio
and carbon excess reflect the occurrence
of the third dredge-up and  HBB in TP-AGB stars with different masses
and metallicities.
Particularly interesting is the case of the M=4~M$_{\odot}$, $Z=0.001$ model, that
 undergoes several crossings through C/O$\,=1$.
See the text for more details.}
\label{fig_agbmodz001}
\end{figure*}
\begin{figure*}
\centering
\includegraphics[angle=90,width=0.95\textwidth]{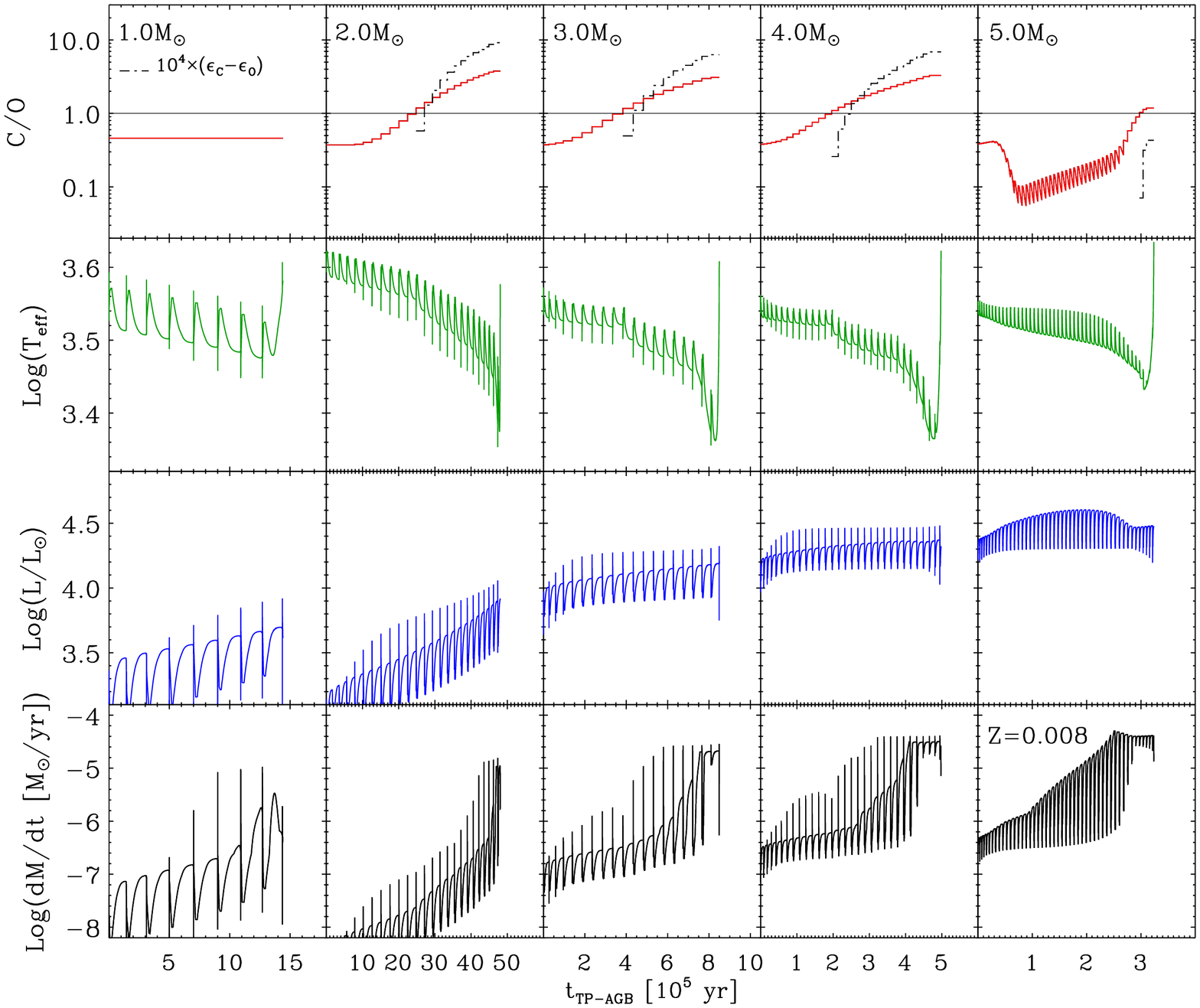}
\caption{The same as in Fig.~\ref{fig_agbmodz001},
 but for initial metallicity $Z=0.008$.}
\label{fig_agbmodz008}
\end{figure*}
\begin{figure*}
\centering
\includegraphics[angle=90,width=0.95\textwidth]{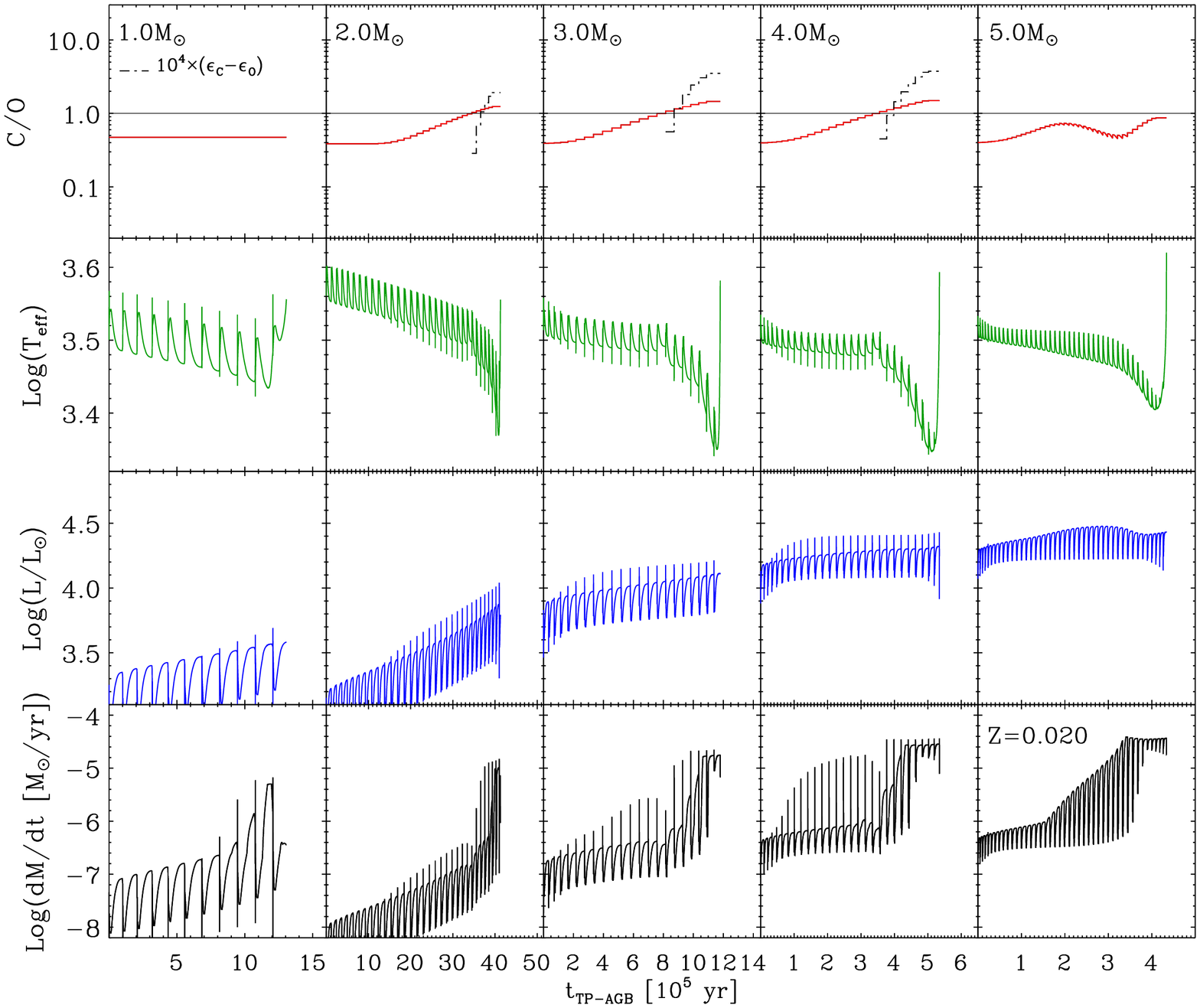}
\caption{The same as in Fig.~\ref{fig_agbmodz001},
 but for initial metallicity $Z=0.02$.}
\label{fig_agbmodz02}
\end{figure*}

Following the results of \citet{marigoetal13}
the evolution of the surface C/O ratio is essentially governed by the
competition between the third dredge-up occurring at thermal pulses,
and HBB taking place during the quiescent interpulse periods. Of great importance
are also the stages of onset and quenching of the two processes,
that normally do not coincide.

The resulting picture is quite complex as
transitions through C/O$=1$ may happen at different stages during the TP-AGB evolution
and with different characteristics.
Figures~\ref{fig_agbmodz001}, \ref{fig_agbmodz008} and
\ref{fig_agbmodz02} show the time evolution of C/O ratio,
effective temperature, luminosity, and mass loss rate
for a few TP-AGB stellar models with selected values
of the initial mass ($M=1,\,2,\,3,\,4,\,5$~M$_{\odot}$) and
metallicity ($Z=0.001,\,0.008\,0.02$).
Together with the C/O ratio we plot also the carbon excess, defined as
the difference between the number  of carbon and oxygen atoms, normalized to
that of hydrogen atoms, $\epsilon_{\rm C}$-$\epsilon_{\rm O}$=$N_{\rm C}/N_{\rm H}$-$N_{\rm O}/N_{\rm H}$.
This quantity is even more meaningful than the C/O ratio,
as  it defines approximately the number of
carbon atoms not locked into the CO molecule, and thus available
for the formation of the carbonaceous dust.
From inspection of Figs.~\ref{fig_agbmodz001}-\ref{fig_agbmodz02} we
extract a few representative cases of the C/O evolution during the
TP-AGB phase, that
are relevant for the interpretation of the results presented in this work.
\begin{itemize}
\item[a)] C/O$\,<1$ during the whole AGB evolution.
This is typical of low-mass stars ($M\approx 1$~M$_{\odot}$),
but the actual minimum mass for a star
to become a carbon star, M$_{\rm C}^{\rm min}$,
is expected to depend critically on
the metallicity, increasing at larger $Z$.
\item[b)] C/O$\,<1 \rightarrow$ C/O$\,>1$ as a consequence of the carbon
enrichment due to the third dredge-up at thermal pulses. This explains
the existence of carbon stars with typical masses of $\approx 2-3$~M$_{\odot}$
depending on metallicity and model details. In these models the rise  of the C/O ratio
begins at the beginning of the third dredge-up
and it is not inhibited by the HBB. Notice that, in these cases,
the excess carbon  becomes larger and larger at decreasing initial metallicity.
Thus at decreasing metallicity, not only the carbon abundance reaches O the abundance
more easily but also the amount of primary carbon production is larger.
The transition to C/O$\,>1$ may be also experienced by the
brightest mass-losing objects (e.g. M> 5 M$_{\odot}$ and $Z=0.008$),
that become carbon rich in their last evolutionary stages
when HBB is extinguished, while few more third dredge-up episodes
may still take place.
\item[c)] C/O$\,<1 \rightarrow$ C/O$\,>1$ as a consequence of a very efficient
HBB, mainly in quite massive AGB stars, $M \ga 5$~M$_{\odot}$,
at very low metallicities, $Z=0.001$.
Differently from the standard channel of carbon star formation, illustrated
in the former examples, in this case the C/O ratio
becomes larger than unity because of the activation of the ON cycle
that causes the destruction of oxygen in favour of nitrogen.
\item[d)] C/O$\,>1 \rightarrow$ C/O$\,<1$  may take place in
relatively massive AGB stars in which
HBB develops after the transition to the carbon star regime.
In these cases the CN cycle  becomes efficient enough to bring the
 star back to the M spectral type (see the crossing of C/O in
the model of 4~M$_{\odot}$ and $Z=0.001$, at time $t_{\rm TP-AGB}\simeq$ 1.2~$\times$~10$^{5}$ yr).
\item[e)] C/O$\,<1 \rightleftarrows$ C/O$\,>1$:  multiple transitions
across  C/O$=1$ may take place under particular conditions.
The sawtooth trend  of the C/O ratio around unity
occurs for a significant part of the TP-AGB evolution of the
M=4~M$_{\odot}$, $Z=0.001$ model. This peculiar
phase  is characterized by quasi-periodic transitions across  C/O$=1$
from both below and above unity, caused by the alternating effects of
the third dredge-up (C/O $\uparrow$) and HBB (C/O $\downarrow$),
so that two crossings may be experienced within a single thermal pulse cycle.
\end{itemize}

The TP-AGB models used in the present work are based on
specific prescriptions for the mass loss
and the third dredge-up, which we briefly outline below.

\noindent{$\bullet$ \em Mass loss.}
It has been included under the hypothesis that it is driven by two main mechanisms,
dominating at different stages. Initially, before radiation
pressure on dust grains becomes the main agent of stellar winds,
mass loss is described with the semi-empirical relation
by \citet{SchroderCuntz_05}, which essentially assumes that the stellar wind originates from
magneto-acoustic waves operating below the stellar chromosphere.
The corresponding mass loss rates are indicated with
$\dot M_{\rm pre-dust}$.

Later on the AGB
the star enters the dust-driven wind regime, which is treated  with an approach
similar to that developed by \citet{Bedijn_88}, and recently adopted by \citet{Girardi_etal10}.

Briefly, assuming that the
wind mechanism is the combined effect of two processes, i.e., radial
pulsation and radiation pressure on the dust grains in the outermost
atmospheric layers, we adopt a functional form of the kind
$\dot{M} \propto \exp({M^{\alpha} R^{\beta}})$,
as a function of the current stellar mass and radius. This expression synthesizes
the results of numerical computations of periodically-shocked atmospheres
\citep{Bedijn_88}. The free parameters $\alpha$ and $\beta$
have been calibrated on a sample
of Galactic long-period variables with measured mass loss rates, pulsation
periods, stellar masses, radii, and effective temperatures.  More details
about the fit procedure will be given elsewhere.
We denote the corresponding mass loss rates with $\dot M_{\rm dust}$.
At any time during the TP-AGB calculations
the actual mass loss rate is taken as the maximum between
$\dot M_{\rm pre-dust}$ and $\dot M_{\rm dust}$.

The key feature of this formalism is that it predicts
an exponential increase of the mass loss rates as the evolution
proceeds along the TP-AGB, until typical
superwind values, around $10^{-5}-10^{-4}$~M$_{\odot}$yr$^{-1}$, are
eventually reached (see bottom panels of Figs.~\ref{fig_agbmodz001}-\ref{fig_agbmodz02}).
The super-wind  mass loss is described in the same fashion as in
\citet{Vassiliadis93},  and corresponds
to a radiation-driven wind, $\dot M_{\rm sw}=L/c\, v_{\rm exp}$,
where $c$ is the speed of light and $v_{\rm exp}$ is the terminal velocity
of the wind.

At any time during the TP-AGB calculations
the actual mass loss rate is taken as
\begin{equation}
 \dot M =
{\rm max}[\dot M_{\rm pre-dust}, {\rm min}(\dot M_{\rm dust},
\dot M_{\rm sw})].
\end{equation}

\noindent{$\bullet$ \em The third dredge-up.}
The onset of the third dredge-up is predicted with the aid of envelope
integrations requiring that a minimum temperature ($\log(T)=6.4$
for the present calculations) should be
reached at the base of the convective envelope.
The rather low value of the temperature parameter favours an earlier
occurrence of the  third dredge-up episodes.
The efficiency of the third dredge-up  is computed with
the analytic fits to full TP-AGB models of
\citet{Karakas_etal02}, as a function of
current stellar mass and metallicity.
The chemical composition of the dredged-up material
(mainly in terms of $^{4}$He, $^{12}$C, $^{16}$O, $^{22}$Ne
and  $^{23}$Na)
is obtained with the aid of a full nuclear network applied to a
model for the pulse-driven convection zone.

\noindent{$\bullet$ \em Hot-bottom burning}.
Both the break-down of the core mass-luminosity relation and
the rich nucleosynthesis of HBB (via the CNO, NeNa and MgAL cycles)
are followed in detail, solving a complete nuclear network
coupled to a diffusive description of mixing, at each mesh
throughout the convective envelope.
The peculiar bell-shaped luminosity evolution
of massive TP-AGB stars with HBB is shown in Figs.~\ref{fig_agbmodz001}-\ref{fig_agbmodz02}.

It should be mentioned that the present set of TP-AGB models is a
preliminary release, since we are currently working to a global TP-AGB
calibration, aimed at reproducing a large number of AGB observables at
the same time (star counts, luminosity functions, C/M ratios,
distributions of colors, pulsation periods, etc.) in different star
clusters and galaxies.  Since the calibration is still ongoing, the
current parameters (e.g. efficiency of the $3^{\rm th}$ dredge-up
and mass loss) of the TP-AGB model may be changed in future
calculations.

Anyhow, various tests indicate that the present version of  TP-AGB models
already yields a fairly good description of
the TP-AGB phase.
 Compared to our previously calibrated sets
\citep{MarigoGirardi_07, Marigo_etal08, Girardi_etal10} the new TP-AGB models
yield somewhat shorter, but still comparable,
TP-AGB lifetimes, and
they successfully recover various observational constraints dealing with
e.g. the Galactic initial-final mass relation,
spectro-interferometric determinations of AGB stellar parameters
\citep{Klotz_etal13}, the correlations
between mass loss rates and pulsation periods, and the trends
of the effective temperature with the C/O ratio observed in
Galactic M, S and C-stars.

As we will show later in this paper, further important support
comes from the results of the dust growth model applied to the
TP-AGB tracks, which are found to nicely reproduce other independent
sets of observations, i.e. the correlation between
expansion velocities and mass loss rates of Galactic AGB stars
(see Section~\ref{subsec_vel}).

\section{Wind model}
\label{sec_wind}
The TP-AGB evolutionary models computed with \texttt{COLIBRI} provide the
input parameters necessary to determine the structure of the expanding
CSE at each time step, namely:  effective temperature ($\Teff$), luminosity (L$_*$),
current mass  (M$_*$),  mass loss rate ($\dot{M}$),  and photospheric
chemical composition.
\subsection{Wind dynamics}
\label{subsec_wind_dynamics}
Following FG06,
 the equations below describe a stationary and spherically symmetric outflow
of one-fluid component, assuming that
there is no drift velocity between gas and dust.
Neglecting the contribution of pressure we have
\begin{equation}\label{dvdr}
 v \frac{dv}{dr}=-\frac{G M_*}{r^2}(1-\Gamma)\,,
\end{equation}
where
\begin{equation}\label{gamma}
 \Gamma=\frac{L_*}{4\pi c G M_*}\, \kappa
\end{equation}
is the ratio between the radiative and the gravitational accelerations.
The opacity $\kappa$,
expressed as mean absorption coefficient per unit mass [cm$^2$g$^{-1}$],
is given by the sum of the gas contribution and of all the dust species
\begin{equation}\label{kh}
 \kappa= \kappa_{gas}+\sum_i f_i \kappa_{i},
\end{equation}
where $\kappa_{gas}=10^{-8}\rho^{2/3} T^3$ \citep{Bell94}, $f_i$ is the degree of condensation
of the key-element\footnote{Following FG06, the key-element of a given
dust species
denotes the least abundant element among those involved in its formation;
e.g., Si is the key-element of the silicate compounds in O-rich stars.}
 into a certain dust species $i$ and $\kappa_i$ is its
opacity,
computed assuming the complete condensation of the key-element initially available
in the gas phase. Once dust is formed,  its opacity increases and
the contribution of the gas becomes negligible. The wind
accelerates if the opacity becomes large
enough so that $\Gamma>1$. As shown in Eq.~(\ref{dvdr}), the
acceleration is higher if this condition is satisfied at short radial distances, as it is proportional to
$r^{-2}$.
The degree of condensation can be written as,
\begin{equation}\label{dcond}
 f_i=n_{k,i}\frac{4\pi (a_i^3-a_0^3)\rho_{d,i}}{3 m_{d,i} \epsilon_{k,i}} \epsilon_s,
\end{equation}
where $n_{k,i}$ is the number of atoms of the key-element present
in one monomer of the dust species $i$;
$m_{d,i}$ is the mass of the monomer;
$a_i$ denotes the grain size (radius) and $a_0$ the initial grain size;
$\rho_{d,i}$ is the dust density of the grain;
$\epsilon_{k,i}$, $\epsilon_s$
are the number densities of the key-element,
and of the initial number of dust grains (seed nuclei)
normalized to the number density of hydrogen $N_{H}$, respectively.

The wind dynamics and its final velocity are determined both
by the amount of dust produced and its chemistry.
Different dust species, in fact, have different optical
properties, hence opacities, so that they
condense at different radial distances.
The most opaque and abundant dust species are amorphous
silicates (olivine and pyroxene) in M-stars, and
amorphous carbon in C-stars, whereas the other species
give a negligible contribution to the total opacity.

The density profile $\rho(r)$ across the wind is determined by the continuity equation:
\begin{equation}\label{rho_pr}
\rho(r)=\frac{\dot{M}}{4\pi r^2 v}\, .
\end{equation}

The temperature structure $T(r)$ is described with
the approximation for a grey and spherically symmetric extended atmosphere
\citep{Lucy71, Lucy76}
\begin{equation}\label{T_pr}
 T(r)^4=\Teff^4 \Big[W(r)+\frac{3}{4}\tau\Big],
\end{equation}
where
\begin{equation}\label{W_r}
 W(r)=\frac{1}{2}\Big[1-\sqrt{1-\Big(\frac{R_*}{r}\Big)^2}\Big]
\end{equation}
represents the dilution factor, $R_{*}$ is the photospheric radius,
and $\tau$ is the optical depth that
obeys the differential equation
\begin{equation}\label{dtaudr}
 \frac{d \tau}{d r}=-\rho \kappa \frac{R_{*}^{2}}{r^2},
\end{equation}
with the boundary condition
\begin{equation}\label{taufin}
 \lim_{r\rightarrow\infty}\tau=0.
\end{equation}

\subsection{Treatment of the dust opacities}\label{sec_opacity}

The opacity of the dust is a critical quantity that
determines the efficiency of the wind, as it enters the momentum conservation
through the Eddington factor in Eq.~(\ref{gamma}).
For each dust species under consideration, the corresponding
opacity as a function of the wavelength
is computed by averaging
the radiation pressure cross-section, $\pi a^2 Q_{\rm rp}(a, \lambda)$,
over the normalized grain size distribution  $d\epsilon/da_i$,
\begin{equation}\label{sigmal}
 \kappa_{i, \lambda}=\frac{\pi}{m_{\rm H}}\int_{a_{\rm min}}^{a_{\rm max}} a^2 Q_{\rm rp}(a, \lambda) \frac{d\epsilon}{da}_i da,
\end{equation}
where $m_{\rm H}$ is the mass of the hydrogen atom,
and $d\epsilon/da_i$ is expressed in units of $\mu$m$^{-1}$.
The radiation pressure cross-section,
$Q_{\rm rp}$, is calculated by means of the Mie theory \citep{Hoyle91}
over a grid of sizes and wavelengths,
assuming the dust grains to be spherical. The quantity $Q_{\rm rp}$ is related to the absorption
and scattering coefficients, $Q_{\rm abs}$ and $Q_{\rm sca}$, by the relation
\begin{equation}
 Q_{\rm rp}=Q_{\rm abs}+Q_{\rm sca}(1-g),
\end{equation}
where $g$ is the asymmetry parameter that describes the deviation from isotropic scattering.
The computation is performed starting
from the optical constant\footnote{In the standard terminology
of solid state physics, the optical constants  ($n$, $k$) of a solid correspond to
the refractive index, and the absorptive index, respectively.} data ($n$, $k$)
by means of a Matlab routine developed by Christian M\"{a}tzler\footnote{The routine
is available at \url{http://omlc.ogi.edu/software/mie/}}.
The opacities of the silicate dust are presently debated.
Since those of  \citet{Dorschner95},
adopted in \citet{GS99}, are found not large enough
to produce a satisfactory terminal velocity of the outflow
 \citep{Jeong03}, both FG06 and \citet{ventura12}
 adopt the opacities from \citet{Ossenkopf92}.
There is another set of data \citep{LeSid96}  which is
more opaque at wavelengths shorter than 9.8~$\mu$m \citep{Jeong03},
but we find that the results obtained adopting this latter set of data
are very similar to those
obtained using the opacities of \citet{Ossenkopf92}.

For this  work the assumed set of dust opacities is as follows.
For M-stars, the ($n$, $k$) data of olivine and pyroxene are taken
from \citet{Ossenkopf92},  and for corundum we refer to \citet{Begemann97}.
For C-stars the ($n$, $k$) data of amorphous carbon are
from \footnote{Taken from http://www.mpia-hd.mpg.de/HJPDOC/carbon.php},  while we use \citet{Pegourie88} for silicon carbide.
Iron opacity is derived from \citet{Leksina67}.

For each dust species $i$, the grain size distribution is assumed to follow a power law:
\begin{equation}\label{dnda}
 \frac{d\epsilon}{da}_i=A_i a^{-x_g},
\end{equation}
where the size $a$ is normalized to 1~$\mu$m,
  $A_i$ is the normalization constant and the slope $x_g$ is the same for all
the dust species.
The quantity $A_i$ is derived for the case of complete condensation
by considering that the total mass of dust that can be formed can be expressed either as
\begin{equation}\label{dust_mass2}
 M_i=\frac{\epsilon_{k,i}}{n_{k,i}} m_{d,i} N_{\rm H},
\end{equation}
or, after integrating over the grain size distribution, as
\begin{equation}\label{dust_mass}
 M_i=\rho_{d,i}N_{\rm H}\int_{a_{\rm min}}^{a_{\rm max}}\frac{4\pi a^3}{3}A_ia^{-x_g} da,
\end{equation}
where $a_{\rm max}$ and $a_{\rm min}$ are respectively the maximum and the minimum dust size of the
distribution.
By equating the two above expressions the normalization constant reads
\begin{equation}\label{A}
 A_i=\frac{3\epsilon_{k,i} m_{d,i} (4-x_g)}{n_{k,i} 4 \pi\rho_{d,i} (a_{\rm max}^{4-x_g}-a_{\rm min}^{4-x_g})}.
\end{equation}
This quantity scales with the abundance of the
key-element (proportional to the metallicity) and so does the opacity in Eq.~(\ref{sigmal}).

For each dust species the opacity is averaged over
the incident radiation field approximated with a black body.
Following \citet{Lamers99} we consider two limiting cases, the optically thin and the optically thick medium.
As long as the medium remains optically thin, we use the Planck average over
a black body at the effective temperature of the star, $\Teff$.

When dust begins to form, the opacity rises by orders of magnitude
and eventually the medium becomes optically thick.
In this case we use the Rosseland mean opacity and the black body distribution
is computed at the local gas temperature, assuming local thermal equilibrium.

To account for the transition from an optically thin to an optically thick medium
we use a linear combination of the Planck and Rosseland opacities,
with a weighting factor given by the ``vertical optical depth", $\tau_d$
\begin{equation}\label{k_av}
 \kappa_{av}=\kappa_{P}\exp(-\tau_{d})+\kappa_{R}[1-\exp(-\tau_d)],
\end{equation}
where the Planck and the Rosseland opacities, $\kappa_{P}$ and $\kappa_{R}$,
are computed summing the contribution of all the dust species as in Eq.~(\ref{kh}).
In the computation of $\tau_d$ we do not consider the dilution factor
because the weighting factor is computed from the condensation radius, $R_{\rm cond}$, outwards, and
dust condensation occurs very rapidly.
\begin{equation}
\label{tau_lin}
 \tau_d = \int_{R_{\rm cond}}^{r} \rho \kappa_{av} dr'.
\end{equation}
The opacity $\kappa_{av}$ is actually used in Eq.~(\ref{gamma}).

\subsection{The dust condensation temperature $T_{\rm cond}$}
\label{ssec_tdust}
Another important quantity for the calculation of a dusty CSE model
is the dust equilibrium temperature.
It is the temperature attained by a dust grain when it reaches
the equilibrium between the energy it absorbs from the radiation field
and the energy it re-emits.
In particular we are interested in the equilibrium temperature, T$_{\rm dust}$, that the dust would have at the condensation point.
Since at this point the medium is optically thin, we can express the energy balance in the following way \citep{Lamers99}
\begin{equation}\label{energy_balance}
 T_{\rm dust}^4 Q_{\rm abs, P}(a, T_{\rm dust})= \Teff^4 Q_{\rm abs, P}(a, \Teff) W(r),
\end{equation}
where $Q_{\rm abs, P}$ is the Planck averaged absorption coefficient expressed explicitly as
a function of the temperature and grain size, and the dilution factor
W(r) is defined in Eq.~(\ref{W_r}).
We notice that this is only a virtual quantity, i.e. it is the temperature that the dust would have if
it could begin to form at a given radial distance.
We define the dust condensation temperature, $T_{\rm cond}$,
as the dust equilibrium temperature at the point where
dust of a given type effectively condenses.
In principle, one should take into account that, if different dust species form at different radial distances,
Eq.~(\ref{energy_balance}) should hold only for species that forms first
because thereafter the medium could be no more optically thin.
However, as we will show below, this approximation is applied only to the silicate dust
produced by  M-giants, that is by far the most important opacity source
in their envelopes and for which we assume a condensation temperature independent of composition
(olivine or pyroxene).
\section{Dust growth}
\label{sec_growth}
Dust formation in the expanding CSE of an AGB star  can be modeled
as a two-step process. Initially, small stable refractory aggregates
are formed from the molecules in the gas phase (seed nuclei),
then, as the temperature decreases,
further accretion on their surface occurs by addition
of more molecules.
Following a commonly accepted scenario,  the first to form are
the most refractory aggregates, and subsequently the process
proceeds by heterogeneous accretion \citep{Gail86, Jeong03}.
However, the details of this process are so poorly understood that,
for example, in the case of M-giants even  the chemical composition of these
aggregates is still a matter of debate \citep{Jeong03}.

In our model,
once the first seeds are formed, further accretion  is described
by explicitly computing the sticking rate of molecules on the grain surface
and their evaporation rate due to the destruction processes.
The primary destruction process is sublimation,
caused by heating of the grains due to absorption of stellar radiation.
Another mechanism is chemisputtering by H$_2$ molecules.
\citet{GS99} and FG06 assume this mechanism to be very efficient
within the CSE of M-type AGB stars.

Below we outline
the basic equations that govern the dust growth model,
while we refer to \citet{GS99} and FG06 for a more detailed description.

\subsection{Seed nuclei}\label{sec_seed}
Following \citet{GS99} we treat the abundance of the seed nuclei, $\epsilon_s$,
as a free parameter.
For silicates, a rough estimate of this quantity can be obtained using Eq.~(\ref{dnda}) assuming
the number of seeds to be equal to the number of grains needed to reproduce the observed
local ISM extinction.
Adopting an exponent $x_g$=3.5,
$\log{A_{\rm ol}}=-15.21$ \citep[value suitable for olivine, ][]{Mathis77},
and integrating between
$a_{\rm min}=0.005$~$\mu$m and $a_{\rm max}=0.25$~$\mu$m,
we get
\begin{equation}
\label{eq_epseed}
 \epsilon_s=\int_{a_{\rm min}}^{a_{\rm max}}\frac{d\epsilon}{da}_ida\sim 10^{-10}.
\end{equation}
With this initial number of seeds we obtain
grain sizes of about $a\sim$0.01~$\mu$m for the M-star models, never
succeeding in reproducing the largest observed grain size $a\ge$~0.1~$\mu$m \citep{Sargent10, Norris12}.
On the other hand, assuming that the seeds correspond only to the fully-grown grains
with typical sizes $a$=0.1~$\mu$m and mass $m_g=(4/3) \pi a^3 \rho_d$  \citep{Jones76, vanloon05},
we can estimate their number from the grains needed to reproduce the total dust mass from Eq.~(\ref{dust_mass2}):
\begin{equation}
 \epsilon_s=\frac{M_i}{m_g}\frac{1}{N_{\rm H}}.
\end{equation}
Assuming for olivine dust
$\rho_d=3.75$ g cm$^{-3}$, $n_{\rm Si, ol}=1$ if
the key-element is silicon and $\epsilon_{\rm Si, ol}=3.55 \cdot 10^{-5}$ according to
the abundances of \citet{Anders89}, we get
$\epsilon_s\sim 10^{-13}$, a much lower value than that of Eq.~(\ref{eq_epseed}).
This abundance of seeds, $\epsilon_s=10^{-13}$, assumed by \citet{GS99},
appears  to be consistent with detailed nucleation computations by \citet{Jeong03},
as well as in very good agreement with the value inferred by
\citet{Knapp85}  for a sample of Galactic M-giants.
We note that adopting  $\epsilon_s=10^{-13}$ in our models, the
grains reach typical sizes of $\sim$0.15~$\mu$m, in agreement with the observations.
Anyhow, it is reasonable to keep $\epsilon_s$ as an adjustable parameter,
since  the details of the nucleation processes and even the composition of the actual seeds
are still very uncertain \citep{Goumans12}.
Similar considerations and  results hold also for amorphous carbon.

At varying metallicity it is natural to expect that the number of seeds
depends also on the abundance of the molecules that form
the initial aggregates. Because of the above uncertainties on
the process of seed formation we assume that
in M-giants the number of seeds scales with the gas metallicity:
\begin{equation}
 \epsilon_{s,M}=\epsilon_s \Big(\frac{Z}{Z_{\rm ISM}}\Big)\,,
\label{nseeds_M}
\end{equation}
where Z$_{\rm ISM}=0.017$ is the local metallicity of the ISM.
 Analogously, for C-stars we scale the number of seeds with
the abundance of carbon not locked into CO molecules,
$\epsilon_{\rm (C-O)}=\epsilon_{\rm C}-\epsilon_{\rm O}$
\citep{Cherchneff06}

\begin{equation}
 \epsilon_{s,C}=\epsilon_s \Big[\frac{\epsilon_{\rm (C-O)}}{\epsilon_{\rm (C-O), ISM}}\Big],
\label{nseeds_C}
\end{equation}
where $\epsilon_{\rm (C-O), ISM}$ is evaluated
from equation (\ref{A}) with
 $\log{A_{\rm C}}=-15.24$  \citep{Mathis77}.

\subsection{Accretion of dust grains}
\label{sec_grain}
For each dust species $i$ the growth of dust grains
is determined by the balance between the rate of effective collisions
of molecules on the grain surface, $J^{\rm gr}_i$, and its decomposition rate, $J^{\rm dec}_i$.
The differential equation describing the dust growth is usually expressed in terms
of the variation of the
grain radius, $a_i$, which is obtained from the variation of the grain volume, $V_i$
\begin{equation}
 \frac{dV_i}{dt}=4 \pi a_i^2 V_{0,i}(J^{\rm gr}_i-J^{\rm dec}_i),
\end{equation}
where $V_{0,i}$ is the volume of the nominal monomer of dust and
 $J^{\rm gr}_i$ and  $J^{\rm dec}_i$ are expressed per unit time and per unit surface.
By differentiating  $V_i=(4/3) \pi a_i^3$ one finally obtains
\begin{equation}\label{dadt}
 \frac{da_i}{dt}=\Big(\frac{da_i}{dt}\Big)^{\rm gr}-\Big(\frac{da_i}{dt}\Big)^{\rm dec}=V_{0,i}(J^{\rm gr}_i-J^{\rm dec}_i),
\end{equation}

\subsubsection{The growth rate}
The growth is assumed to proceed through the addition of molecules from the gas phase,
via suitable chemical reactions.
For example forsterite, Mg$_2$SiO$_4$, is assumed to grow
via the following net reaction,
\begin{equation}\label{reaction}
 \rm {2Mg +  SiO + 3H_2O\rightarrow
Mg_2SiO_4(s) + 3H_2,}
\end{equation}
The growth rate for each dust species, $i$, is defined as the minimum between the
rates of effective collisions on the grain surface
of the gas molecules involved in the formation of the dust monomer, through formation
reactions like that in Eq.~(\ref{reaction}).
The molecular species determining such a rate  is named ``rate-determining species''.
\begin{equation}\label{growth}
  J^{\rm gr}_i=\min\Big [s_i\frac{\alpha_i n_{j, g} v_{th,j}(T_{gas})}{s_j}\Big ],
\end{equation}
where $n_{j, g}$ is the number density of the molecule $j$ in
the gas phase, $T_{\rm gas}$ is the gas temperature given by Eq.~(\ref{T_pr}), $v_{th, j}(T_{\rm gas})$
the corresponding thermal velocity, $s_j$ its stoichiometric
coefficient in the dust formation reaction, $s_i$ the stoichiometric coefficient of the monomer of the dust species $i$ and $\alpha_i$ is its sticking coefficient.

\subsubsection{The destruction rate}
The solid's evaporation proceeds through pure sublimation, $J^{\rm sub}_i$,  due to
heating by stellar radiation and, for dust species that can react with
H$_2$ molecules, such as forsterite,
through the inverse reaction of the analogous of Eq.~(\ref{reaction}) at the grain surface, $J^{\rm cs}_i$ \citep{Helling06}.
The latter process is sometimes named chemisputtering \citep{GS99}.
The total destruction rate is thus
\begin{equation}\label{decomposition}
J^{\rm dec}_i=J^{\rm sub}_i+J^{\rm cs}_i
\end{equation}

$J^{\rm sub}_i$ is determined by
considering that, in chemical equilibrium conditions,
it must equal the growth rate. Since
the sublimation process depends only on the specific properties of
the grain under consideration, the rate so determined must
hold also outside equilibrium.
We thus obtain from Eq.~(\ref{growth}), after eliminating $n_{j, g}$ with the partial pressure and the temperature,
\begin{equation}
\label{dust_subl}
J^{\rm sub}_i= \alpha_i v_{th, j}(T_{\rm dust}) \frac{p(T_{\rm dust})}{k_B T_{\rm dust}},
\end{equation}
where $T_{\rm dust}$ is the dust equilibrium temperature
determined with Eq.~(\ref{energy_balance}), $v_{th, j}(T_{\rm dust})$
is the thermal velocity of the molecule ejected from the grain surface and $k_B$ is the Boltzmann constant.
The quantity  $p(T_{\rm dust})$ is the saturated vapour pressure at the dust temperature,
that can be expressed with the Clausius-Clapeyron equation
\begin{equation}
\label{ptd}
\log p(T_{\rm dust})=-\frac{c_1}{T_{\rm dust}} + c_2
\end{equation}
where $c_1$ and $c_2$ are sublimation constants, characteristic of the species
under consideration.
The constant $c_1$ contains the latent heat of sublimation of the dust species
and the constant $c_2$ actually is slightly dependent on the temperature.
The quantities $c_1$ and $c_2$ may be obtained either directly from
thermodynamical data \citep{Duschl96},
or by fitting with Eq.~(\ref{ptd}) the results
of sublimation experiments \citep{Kimura02,Kobayashi09,Kobayashi11}.

$J^{\rm cs}_i$ is determined in an analogous way, assuming that the growth and destruction rate at equilibrium must balance,
and  using the Eq.~(\ref{reaction}) in chemical equilibrium to determine the partial equilibrium pressure of
the rate-determining species, $p_{j, {\rm eq}}$.
We thus obtain
 \begin{equation}\label{Jdec}
J^{\rm cs}_i= s_i \frac{\alpha_i v_{th,j}(T_{\rm gas})}{s_j} \frac{p_{j, {\rm eq}}}{k_B T_{\rm gas}}.
\end{equation}
where $j$ is the rate-determining species
of each dust species $i$,
and  $p_{j, {\rm eq}}$ must satisfy the equilibrium condition for each formation reaction.
For the reaction of formation of forsterite (\ref{reaction}), the relation is
\begin{equation}\label{partial}
\frac{p^3_{ \rm {H_2}}}
{p^2_{\rm {Mg}} p_{ \rm {SiO}} p^3_{ \rm {H_2O}}}= \exp\Big(-\frac{\Delta G}{RT_{\rm gas}}\Big),
\end{equation}
where the partial pressures are normalized to the standard reference pressure at 1 bar, $\Delta G$ is the variation of the free enthalpy per mole
of the reaction in Eq.~(\ref{reaction})  and $R$ the ideal gas constant.
If, for example, the rate-determining species of forsterite is SiO,  Eq.~(\ref{partial}) must be inverted to find its equilibrium pressure $p_{\rm SiO, eq}$,
given the gas partial pressures of the other species.
Equations (\ref{dadt}),  (\ref{growth}) and (\ref{decomposition}) must be written for each of the dust species considered,
as described in detail in FG06.

\subsection{Low-condensation temperature models}
\label{sec_chemi}
We compute a first set of models
strictly following the methodology outlined in \citet{GS99} and FG06.

\subsubsection{M-star models with efficient chemisputtering}
From the analysis of experimental results by \citet{Nagahara96}, showing
that the destruction rate of forsterite in presence of H$_2$ molecules
grows linearly with the
gas pressure, \citet{GS99} concluded that
the chemisputtering process should be fully efficient
within the CSEs of AGB stars. They have shown that, in this case,
the destruction rate by chemisputtering is always
much larger than that of sublimation, so that one can
assume that  $J^{\rm dec}_i=J^{\rm cs}_i$.
We assume that chemisputtering is fully efficient
for the species that can react with hydrogen molecules.
For the species that do not react with hydrogen molecules, such as
iron, we consider only the sublimation rate.

\subsubsection{C-star models}
\label{cstarmodels}
For amorphous carbon, we take into account that its growth can
initially proceed through complex reactions of C$_2$H$_2$ addition,
forming  isolated chains that subsequently coalesce into larger cores.
Further growth of carbon mantles on these initial seeds can continue
through vapor condensation \citep{GS99}.
This homogeneous accretion is consistent with the microanalyses of pre-solar graphitic spherules  extracted from the meteorites,
revealing the presence of  nanocrystalline carbon cores
consisting of randomly oriented graphene sheets, from
PAH-sized units up to sheets 3-4~nm in diameter \citep{Bernatowicz96}.
According to \citet{Cherchneff92}, the chain of C$_2$H$_2$ addition reactions
have a bottleneck in the formation of the benzene because it becomes effective
only when the {\it gas}  temperature is below $T_{\rm gas}$=1100~K.
Therefore,  while the sublimation temperature of solid carbon
can exceed $\sim$1600~K, its growth should be inhibited above $T_{\rm gas}$=1100~K.
Thus, following FG06, we do not consider
any destruction reaction in the case of amorphous carbon,
but we assume that it can grow only when $T_{\rm gas}\leq$1100~K.

Finally, we notice that a significant fraction of the pre-solar graphitic spherules
contains internal crystals of metal carbides with composition from nearly
pure TiC to nearly pure Zr-Mo carbide \citep{Bernatowicz96}, but only
rarely SiC \citep{Hynes07}.  These carbides might have served as heterogeneous nucleation centers for condensation of carbon,
opening an alternative path with respect to the homogeneous accretion described above \citep{Croat05}.
This alternative path and its implications for the efficiency of carbon dust growth at varying metallicity
will be discussed in a forthcoming paper.

The results of the present sets of calculations will be discussed in Section~\ref{sec_subli}.
Here we anticipate that the application of
Eq.~(\ref{dadt}) in typical conditions of a CSE is that
silicate dust destruction is very efficient above
$T_{\rm gas}\sim$1100~K  \citep{GS99, Helling06, Gail10}.
We notice that this latter value of the temperature is
improperly referred by some authors as the condensation temperature of silicates,
though the corresponding dust temperature of silicates
can be significantly less than the above value, $T_{\rm cond}\sim$900~K.
In analogy to the case of M-star models with efficient
chemisputtering, we will also refer to these
C-star models  as \textit{low-condensation temperature (LCT) models}.

\subsection{High-condensation temperature models}
\label{sec_subli}
\subsubsection{M-star models with inhibited chemisputtering}
The efficiency of chemisputtering is still a matter of debate (Duschl et al. 1996; Nagahara et al. 2009b;
Tielens, private communication).
On the theoretical side, the main argument is that  chemisputtering  could be inhibited by the high activation
energy barrier of the reduction reaction of silicate dust compounds by H$_2$ \citep{Gardner74, Tso82, Massieon93}.
On the observational side
there is evidence, from  recent experiments by \citet{nagaharaetal09}, that
the condensation temperature of crystalline forsterite (iron-free olivine) and crystalline
enstatite (iron-free pyroxene),
at pressure conditions corresponding to stellar winds,
is between 1400--1500~K, while it is only slightly lower for amorphous silicates ($\sim$1350~K).
Therefore,  the condensation temperatures obtained in the
laboratory are significantly higher than those predicted by
models that include a fully efficient chemisputtering. We notice that these recent results are not in contradiction with the previous
work of \citet{Nagahara96} on the evaporation rate of forsterite,
because the evidence of chemisputtering in their experiment appears at a gas pressure which is considerably higher than
that at which dust is predicted to form in the  CSEs.

Following the above indications, we also compute a class of models without chemisputtering.
The evolution of the grain size
is described by Eq.~(\ref{dadt}), with the destruction rate given only by the sublimation term,
$J^{\rm dec}_i=J^{\rm sub}_i$.
We notice however that the sublimation rate depends on the
dust equilibrium temperature $T_{\rm dust}$ which, in our simple model,
can be defined only near the condensation point (Section \ref{ssec_tdust}).
Thus, instead of integrating the full Eq.~(\ref{dadt}),
we first determine the condensation point within the CSE by comparing the
growth rate with the maximum sublimation rate,
i.e. the sublimation rate obtained by setting $\alpha_i=1$ in Eq.~(\ref{dust_subl}).
This point is defined by the condition $J^{\rm gr}_i = J^{\rm dec}_{i, {\rm max} }$ and provides also
the condensation temperature.
Beyond this point,  we compute the evolution of the grain size
by assuming that the sublimation process is negligible
in the right-hand side of Eq.~(\ref{dadt}).
In this way, the condensation temperature
depends, on one side, on the dust species which determines $J^{\rm dec}_{i, {\rm max}}$ and, on the other,
on the physical conditions of the CSE, which determine $J^{\rm gr}_i$.
Since the real sublimation rate is only $\alpha_i$ times the maximum sublimation rate, the above condition
implies that condensation begins when
the growth rate is $\sim$ $1/\alpha_i$ ($\sim$10 for silicates) the
real sublimation rate. The corresponding super-saturation ratio
is also $\sim$1/$\alpha_i$ ($\sim$10 for silicates).
Considering also that, beyond the condensation point,
the sublimation rate decreases almost exponentially with the
temperature, retaining only the growth term
in Eq.~(\ref{dadt}) does not affect the  accuracy of the results.
This method is similar to the procedure usually followed in the literature,
but for the fact that,  instead of assuming a fixed condensation temperature, we derive
it from the comparison of the growth and destruction rates.

\begin{figure}
 \resizebox{0.9\hsize}{!}{\includegraphics[angle=0]{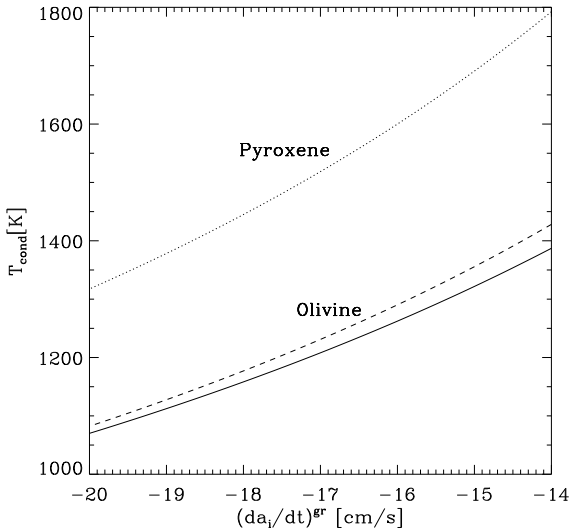}}
        \caption{Condensation temperatures of olivine and pyroxene as a function of the growth rate given by
Eq.~(\ref{dadt}) with $J^{\rm dec}_i=0$. The curves for olivine (solid and dashed lines) are obtained from
the models of \citet{Kimura02} and \citet{Duschl96} respectively. Pyroxene curve is from \citet{Kimura02}.}
        \label{dmdt_dust}
        \end{figure}

An estimate of the condensation temperature of silicates as a function of  $(da_i/dt)^{\rm gr}$,
is shown in Fig.~\ref{dmdt_dust}.
The solid and dotted lines refer to the case of the olivine
and pyroxene, respectively, with  $J^{\rm sub}_i$ computed
following  the method by \citet{Kimura02} (see also  \citet{Kobayashi11}).
For the evaporation of pyroxene they have considered that the preferential
mode is through SiO$_2$ molecules \citep{Tachibana02} and have
empirically derived  $c_1=$6.99~$\times$~10$^{4}$~K and
$\exp(c_2)$=3.13~$\times$~10$^{11}$~dyne~cm$^{-2}$.
With these values in the equation of the saturated vapour pressure (Eq.~\ref{ptd}),
we obtain condensation temperatures that lie  between 1400~K and 2000~K.
For olivine, \citet{Kimura02} determine, from experimental results,
$c_1=6.56~\times~10^{4}$~K and $\exp(c_2)=6.72~\times~10^{14}$~dyne~cm$^{-2}$.
With these values we obtain condensation temperatures between 1200~K and 1400~K.
The olivine curve obtained with this approach can be compared with that
derived for forsterite (dashed line) using, instead, the analytical fits to theoretical calculations
of $c_1$ and $c_2$ provided by \citet{Duschl96}.
We see  that the two methods give
comparable condensation temperatures
for  olivine-type silicates.
These condensation temperatures are significantly
higher than those found in presence of chemisputtering ($T_{\rm cond}\sim$ 900~K).
The values shown in the figure are also consistent with
those derived by \citet{Kobayashi09, Kobayashi11}, in the different context of sublimation
of dust grains in comets.

Moreover, for olivine,  this result is in very good agreement with the recent
experimental results by \citet{nagaharaetal09},
who have found $T_{\rm cond}$ from
$\sim$1350~K (amorphous silicates) to $T_{\rm cond}\sim$1450~K
(crystalline silicates).
Instead, for pyroxene, our values are significantly higher than the above values.
However our result for pyroxene is not very reliable and must be considered only
as un upper limit for the following reason.
We assume that pyroxene evaporates preferentially incongruently through SiO$_2$
but  we do not  take into account that
this incongruent sublimation is followed by
the production of forsterite \citep{Tachibana02} which,
at these temperatures, immediately evaporates.
Thus, in the following,  we will consider
the condensation temperature of  the  olivine
as representative of all other silicates.

We stress that in these models we neglect the chemisputtering process
at any pressure \citep{Kobayashi11} while \citet{Nagahara96} have found that, at
high pressures, the evaporation at fixed temperature increases almost linearly
with the pressure, so that chemisputtering could become efficient.
The lowest pressure in the \citet{Nagahara96} experiment was P$\sim$ $10^{-3}$dyne~cm$^{-2}$
while in \citet{nagaharaetal09} the upper limit was P$\sim$ $10^{-2}$dyne~cm$^{-2}$.
With these two values we derive  typical dust equilibrium temperatures
of T$_{\rm dust}\sim$~1500~K to T$_{\rm dust}\sim$~1600~K, which are
significantly higher than the values of the condensation temperature of
olivine.

Therefore, we conclude that, besides leading
to temperatures in good agreement with
experimental evidence, the assumption of neglecting the
chemisputtering process does not lead to results in disagreement with
the evidence that this process may play a role at higher gas pressures.

\subsubsection{C-star models with modified condensation temperature}
For the amorphous carbon we have already seen  that
the growth is not regulated by its sublimation temperature ($\sim$1500~K),
but by the gas temperature window (900--1100~K) that allows
an efficient chain of  C$_2$H$_2$ addition reactions \citep{Cherchneff92}.
However, recent hydrodynamical investigations indicate that,
during a pulsation cycle,
gas that is initially at temperatures well above $1100$~K,
after being shocked may cool down and remain inside the effective temperature window long enough
to enhance the rate of addition reactions \citep{cherchneff11}.
In the detailed model of IRC+10216 this process allows the formation of amorphous carbon
in inner regions of the CSE where the pre-shock gas temperature is
significantly above the effective window, $T_{\rm gas}=1300$~K \citep{cherchneff12}.
As a consequence, the growth rate of amorphous carbon can be significantly enhanced.
To investigate the impact  on the predicted ejecta of carbon dust,
we explore the case of a higher temperature window,
following the detailed results of \citet{cherchneff12}.
In this case we simply assume that the amorphous carbon can condense
below $T_{\rm gas}=1300$~K, which should set a fairly upper limit
to the growth rate of amorphous carbon.

In the following we will refer this set of models as  \textit{ high-condensation temperature (HCT) models}.
\begin{figure}
\resizebox{0.9\hsize}{!}{\includegraphics{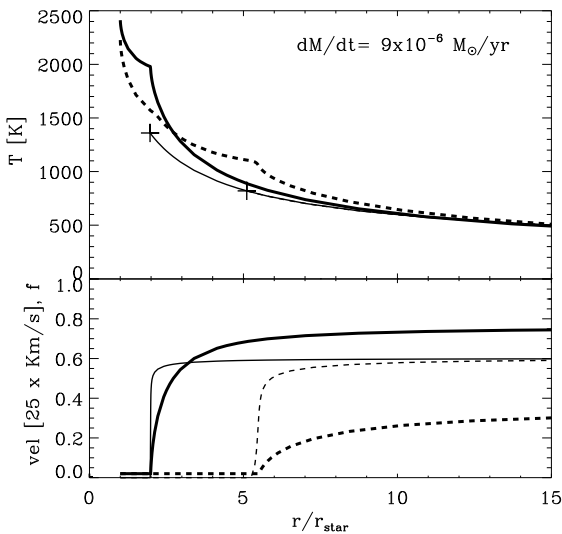}}
 \caption{\textit{Upper panel}: run of gas temperature (thick lines)
 and dust temperature (thin lines) within the CSE for a LCT (dashed) and a HCT (solid) M-star model.
 \textit{Lower panel}: run of velocity (thick lines)
 and total condensation fraction of silicates (thin lines) within a single CSE for LCT model (dashed)
 and HCT model (solid).
The model is for an initial stellar mass of 1.7~M$_{\odot}$, $Z=0.02$,
$\dot{M}\sim 9\times 10^{-6}$~M$_{\odot}$yr$^{-1}$
and with actual star mass of 0.77~M$_{\odot}$, L$_*\sim$ 6.85$\times$~10$^3$ L$_{\odot}$ and $C/O\sim$0.54.
Crosses indicate the beginning of silicate dust condensation in the two models.}
 \label{Dust_temp_vel}
\end{figure}

\subsection{Method of solution and initial conditions}
The system of differential equations describing the dust evolution is given by Eq.s~(\ref{dvdr}), (\ref{dtaudr}) and
an equation like (\ref{dadt}) for each dust species.
In M-stars we consider the evolution of
corundum (Al$_2$O$_3$), quartz (SiO$_2$), iron,
olivine (Mg$_{2x}$Fe$_{2(1-x)}$SiO$_4$)
and pyroxene (Mg$_{x}$Fe$_{1-x}$SiO$_3$).
The quantity $x$, the fractional abundance of Mg molecules with respect to
the sum  of Mg and Fe molecules, ranges from 0 to 1 and
is governed by equations like (12) and (16) of FG06.
In C-stars we consider amorphous carbon, silicon carbide (SiC) and iron.
Finally we add the Eq.~(\ref{tau_lin}) to compute the weighting factor for the opacity.
The independent variable of the system is the radial distance $r$, whereas, the dependent variables are the velocity, $v$, the optical depth,
$\tau$, the grain size for each dust species, $a_i$,  $x$ and the vertical optical depth, $\tau_d$.
The quantities that determine the wind structure are the gas temperature and density profiles, $T(r)$ and $\rho(r)$,
the total opacity, $\kappa$, the dust condensation fraction, $f$, and $\Gamma$.

Since the boundary condition on $\tau$ (Eq.~\ref{taufin}) is at infinity, but
all the other conditions are at the inner radius, we
solve the system by means of the following shooting method.
We begin the integration with a guess value of $\tau_{i}$ at a suitable
inner radius $r_{i}\geq{R_*}$ chosen so that the gas temperature is high enough to inhibit the formation
of any type of dust.
We then integrate outward all the equations until $r=$1000~$R_*$.
The latter condition is set because, at these large distances from the photosphere,
the density is so low that dust cannot grow any more and the acceleration term is negligible.
As the integration moves outward, all the dependent variables, but  $\tau$,
are reset to their initial values, until dust begins to condense.
In this way we avoid that, for example,
the velocity drops to zero before dust condenses, which corresponds to
the assumption that pulsations are able to drive the medium into the condensation region.
We iterate on $\tau_{i}$ until the condition $\tau\leq\epsilon\times\tau_{i}$
is met. This condition actually replaces Eq.~(\ref{taufin})
and the accuracy has been chosen to be $\epsilon=10^{-4}$ in order to provide stable solutions and
to avoid excessive computational time, at once.

The reactions assumed for dust formation are taken from \citet{GS98}, for corundum, and FG06 for all the other dust compounds.
The thermodynamical data for the decomposition rates are taken from \citet{Sharp90}
with the exception of FeSiO$_3$ and SiC for which we use the data taken by \citet{Barin95}.
The values of the sticking coefficients are taken from
laboratory measurements, when
available, or theoretical computations.
However, their uncertainties can be high because they may involve a temperature dependent
activation factor while experiments encompass only a limited range of temperatures.
For olivine we adopt $\alpha_{\rm ol}\sim$ $0.1$ determined from
evaporation experiments of forsterite by \citep{Nagahara96}.
For iron and quartz the experimental values are $\alpha_{\rm Fe}\sim1$ and
$\alpha_{\rm qu}\sim0.01$, respectively \citep{Landolt68}.
For the growth of SiC a value of $\alpha_{\rm SiC}\sim 1$ has
been determined by \citet{Raback99}.
For other species we cannot rely on experimental measurements
of the sticking coefficient and we need to make some assumption.
For pyroxene  we adopt the same value as that used for olivine.
For corundum we chose the maximum possible value, i.e. $\alpha_{\rm co}=1$.
Finally, for amorphous carbon the usual assumption is to adopt $\alpha_{\rm C}=1$ because,
once the C$_2$H$_2$ addition reactions are activated,
the gas temperature $T_{\rm gas}\sim1100$~K is so much below the typical
sublimation temperature $\sim1600$~K of carbon, that the latter is supposed to grow
at the maximum efficiency \citep{Gail88, Cherchneff92, Krueger96}.
We thus adopt the standard assumption  $\alpha_{\rm C}=1$, but we also check
the effects of adopting a lower sticking coefficient  $\alpha_{\rm C}=0.5$ and  $\alpha_{\rm C}=0.1$.

For the computation of the opacities in LCT models, we use the dust distribution parameters
determined by \citet{Mathis77} namely, $x_g$=3.5 and $a_{\rm min}=0.005$~$\mu$m, $a_{\rm max}=0.25$~$\mu$m.
These parameters are kept fixed at varying metallicity. This choice of the parameters
is adopted only for comparison with the results of
FG06 and  \citet{ventura12}.
A different choice is adopted in the HCT models, for which an almost flat distribution is assumed ($x_g$=0.1) and $a_{\rm min}=0.005$~$\mu$m, $a_{\rm max}=0.18$~$\mu$m.
This choice is more consistent  with the grain size distribution obtained from the integration.
The number of initial seeds is set to $\epsilon_s=10^{-13}$ (see also Section~\ref{sec_seed}) and their initial size is assumed to be $a=0.001$~$\mu$m (FG06).

The value of initial velocity of the wind was arbitrary chosen to be 0.5~km~s$^{-1}$.
According to the amount and the dust species produced,
the matter can
a) not being accelerated at all,
b) first accelerate, but then decelerate again at some radial distance
c) be accelerated away from the star.
In any case if the local velocity never exceeds the local escape velocity,
dust is not able to drive the wind -inefficient dust-driven case according to FG06, \citet{Hofner09}-
and we assume that the outflow proceeds at least at the initial velocity.
This assumption tends to overestimate the fraction of dust that
can condense but, since it corresponds to phases where the
mass loss rate is relatively low, it does not significantly affect the total dust
ejecta.

We exclude from the calculations the models for mass loss rates
below 10$^{-8}$ M$_\odot$ yr$^{-1}$, because
dust formation is already negligible at those values (FG06).
\begin{figure*}
\includegraphics[width=0.4\textwidth]{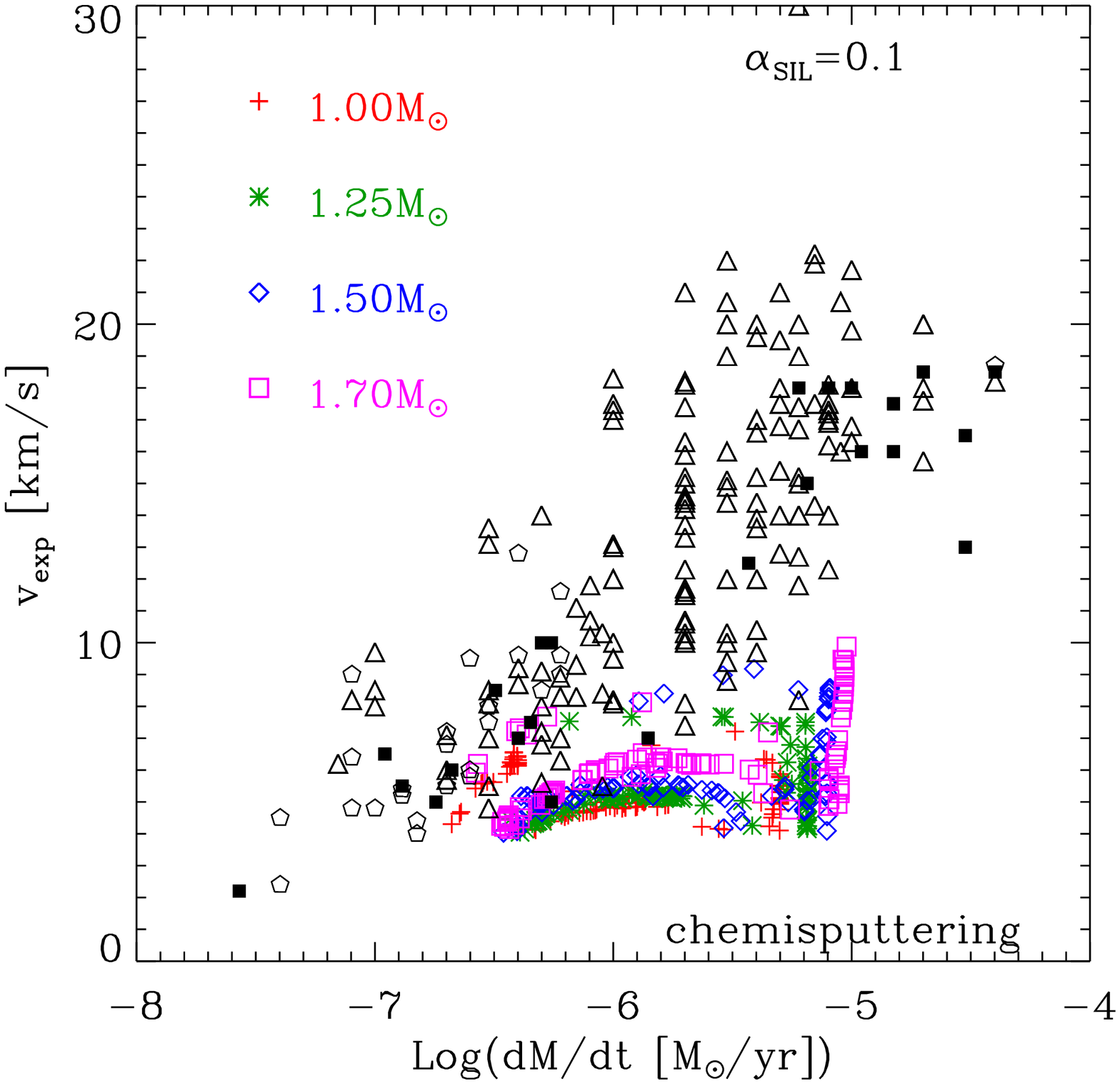}
\qquad \qquad \qquad
\includegraphics[width=0.4\textwidth]{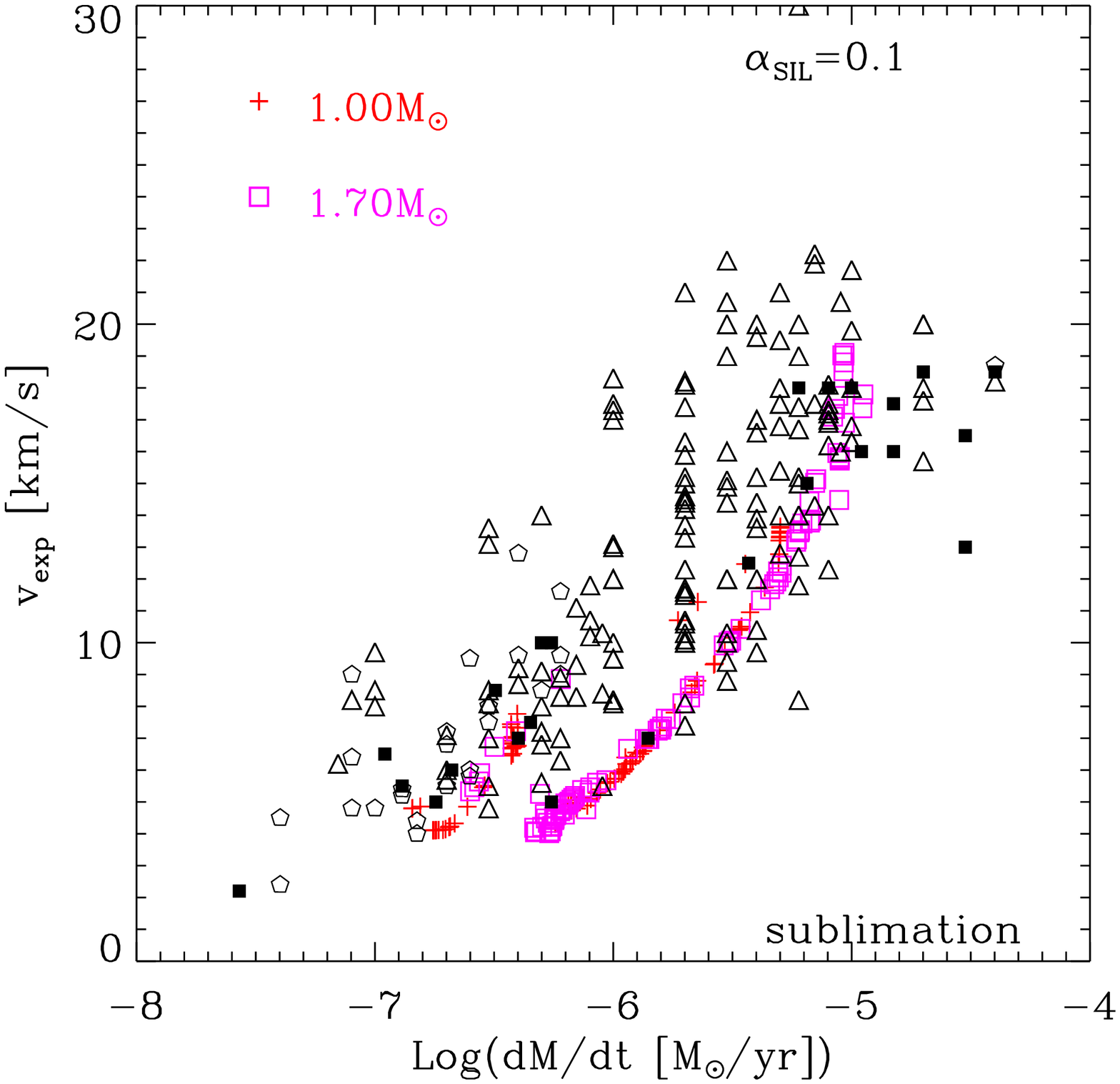}
\caption{Expansion velocities of circumstellar outflows against mass loss rates
of variable M-stars. Observations of Galactic M-stars by \citet{Loup93} (black triangles),
\citet{Gonzalez03} (black pentagons) and \citet{Schoier13} (black squares)
are compared with predicted expansion velocities for a few selected TP-AGB tracks with Z = 0.02
for the values of initial stellar masses listed in upper left of each figure.
\textit{Left panel}: comparison with simulations that assume  fully efficient chemisputtering.
\textit{Right panel}: comparison with HCT models.
}
\label{v_dotm_M}
\end{figure*}

\section{Results}
\label{sec_res}
We follow the growth and evolution of dust in the CSEs of a few selected evolutionary TP-AGB tracks, extracted from the set of \citet{marigoetal13}.

We consider three values of the initial
metallicity, $Z=0.001$, $0.008$, $0.02$, which are representative of low, intermediate and solar
metallicity respectively, and a few selected values of the initial stellar mass,
between 1~M$_{\odot}$ and 6~M$_{\odot}$ as listed in Table~\ref{Table:ejecta}.
We are presently extending the dust computations to other metallicity sets of
TP-AGB models, while the range of Super AGB stars \citep{Siess10}
will be the subject of a following investigation.

\subsection{Expansion velocities}\label{subsec_vel}
\subsubsection{M-stars}
We first discuss the effects of including (LCT) or neglecting (HCT) chemisputtering in
a model representative of
a typical Galactic M-giant, with mass $M=1.7$~M$_{\odot}$, metallicity $Z=0.02$ and
 mass loss rate $\dot{M}\sim 9\times 10^{-6}$~M$_{\odot}$yr$^{-1}$.
In the upper panel of Fig.~\ref{Dust_temp_vel} we show the gas (thick line)  and dust (thin line)
temperature profiles.
In the LCT model dust condensation starts at a {\it gas} temperature of about $T_{\rm gas}\sim$1100~K,
in agreement with \citet{Helling06} and with the condensation curves drawn by \citet{GS99}
for the chemisputtering process. The corresponding dust condensation temperature,
indicated by the lower cross, is only of about $T_{\rm cond}\sim 800$~K.
Instead, when only sublimation is considered, silicate dust
begins to condense at a dust temperature of $T_{\rm cond}\sim 1350$~K,
well above the limit obtained with chemisputtering and
in very good agreement with the one experimentally determined by \citet{Nagahara09}.
We notice that the corresponding gas temperature,
$T_{\rm gas}\sim 2000$~K, is comparable with the
one at which \citet{Nagahara96} performed
their experiments. The corresponding gas pressure is
relatively low, $P\sim 5\times 10^{-2}$ dyne~cm$^{-2}$,  and falls within the regime
where the decomposition rate measured by \citet{Nagahara96} is independent of the pressure.

The velocity and condensation profiles are shown in the lower panel of Fig.~\ref{Dust_temp_vel}.
Notice that the silicate condensation fractions reach
the same value, but the bulk of condensation occurs much closer to the star
when the chemisputtering process is neglected.
In this case, the bulk of condensation happens at $r \sim 2 R_*$,
a factor 2.5 less than when chemisputtering is included ($r \sim R_*$).
The overall effect is that, for the HCT models
the acceleration term is larger, mainly because of the larger local
acceleration of gravity in the inner regions.
This gives rise to a final velocity which is more than twice the terminal velocity reached
by the model with chemisputtering.

We now compare the expansion velocities obtained with our models with those observed
in Galactic M-type AGB stars.
It is important to stress here that, contrary to more sophisticated hydrodynamical models,
our model cannot  provide the mass loss rate, which must be assumed.
The meaning of this comparison is thus to check whether, given the stellar parameters and
the corresponding mass loss rates, our model is able to reproduce the terminal velocities of the wind,
over the entire range of observational data.
The comparison is shown in Fig.~\ref{v_dotm_M}.
Mass loss rates and expansion velocities
are taken from \citet{Loup93} (black triangles),
\citet{Gonzalez03} (black pentagons) and \citet{Schoier13} (black squares).
Data from \citet{Loup93} and \citet{Schoier13} are derived from
observations of $^{12}$CO and HCN rotational transitions, whereas
\citet{Gonzalez03} obtained their values from the interpretation of SiO rotational transition
lines.
The mass loss rates range from $\sim$10$^{-7}$~M$_\odot$yr$^{-1}$
to a few 10$^{-5}$~M$_\odot$yr$^{-1}$, including
also dust-enshrouded AGB stars, while the wind velocities
range from a few km~s$^{-1}$ to 20~km~s$^{-1}$.
For the models we use an initial metallicity $Z=0.02$,
that we consider typical of the solar environment \citep{Lambert86}.
We remind that, though the current solar metallicity is estimated to be $Z_\odot\approx$0.0154 \citep{Caffau_etal11}
and its derived initial metallicity is $Z_\odot\approx$0.017 \citep{Bressanetal12},
\citet{Lambert86} compared their observations with model atmospheres based
on the old \citep{Lambert78}
solar abundances for which $Z_\odot\approx 0.021$.

\begin{figure}
\centering
\includegraphics[width=0.4\textwidth]{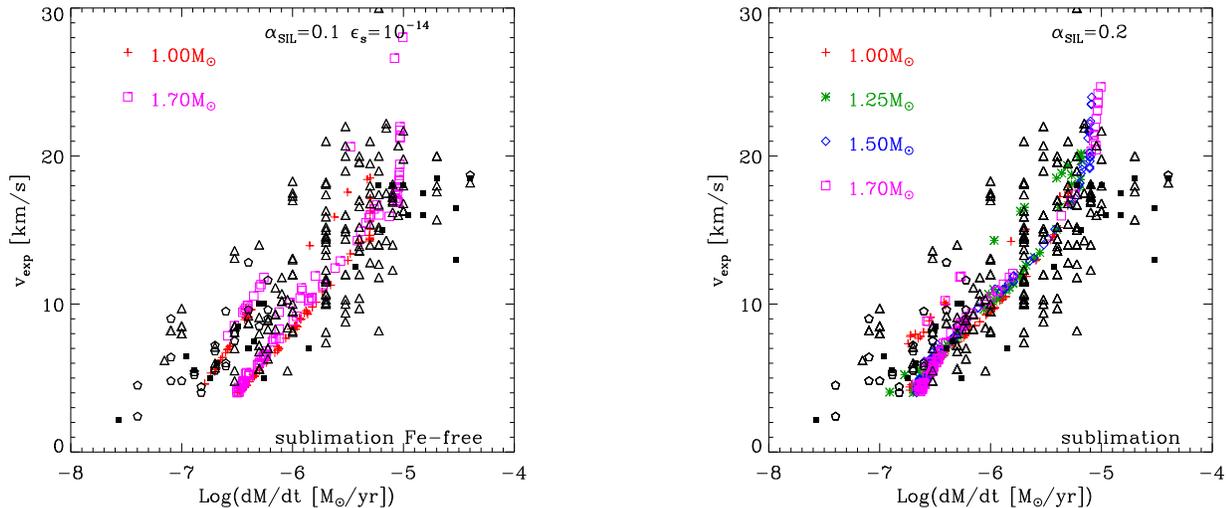}
\caption{The same as in Fig.~\ref{v_dotm_M},
but using HCT models and large iron-free grains ($\sim$0.3~$\mu$m) obtained by decreasing the number of seeds
to $\epsilon_s$=10$^{-14}$.
}
\label{v_dotm_large}
\end{figure}

\begin{figure}
\centering
\includegraphics[width=0.4\textwidth]{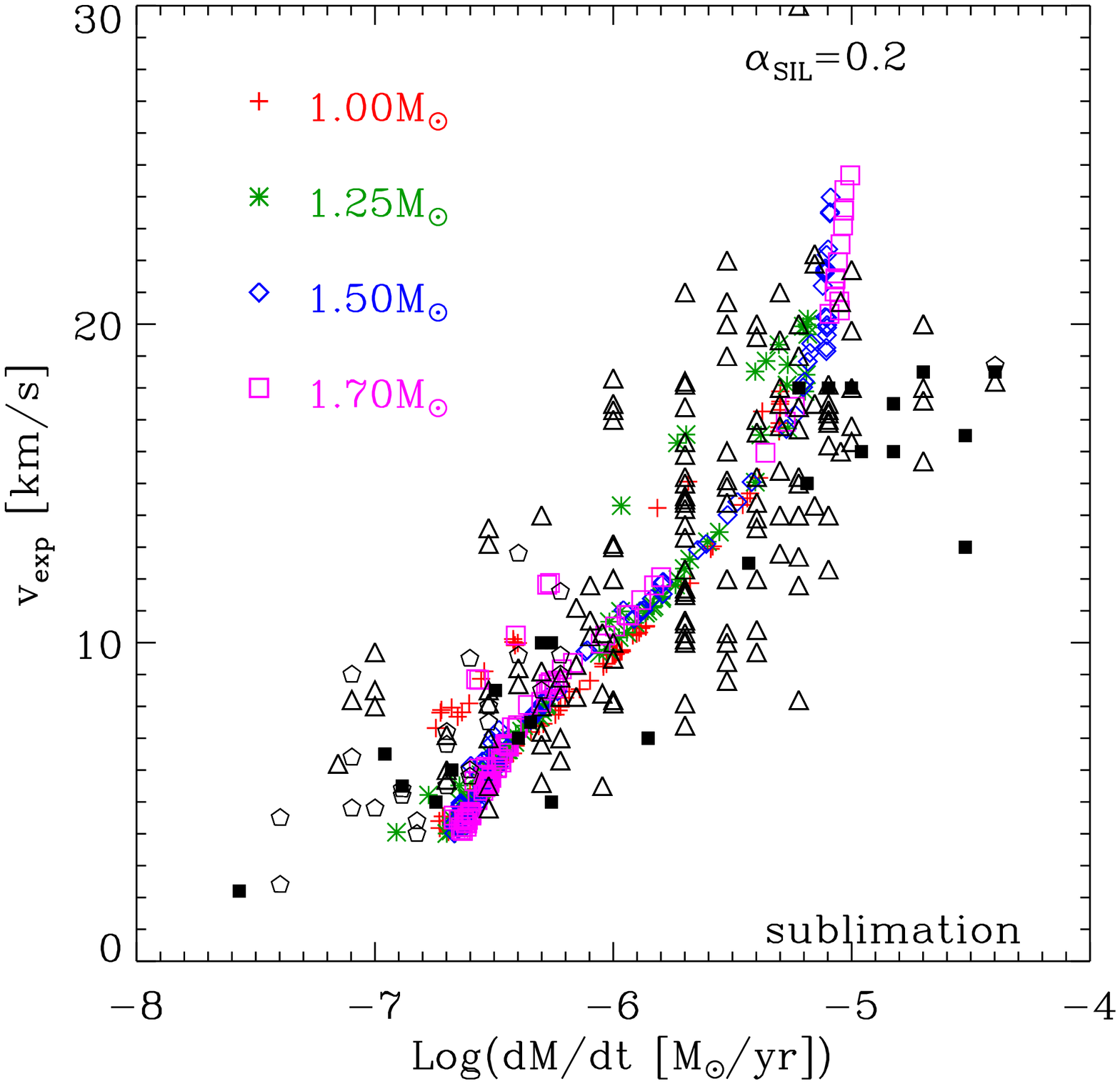}
\caption{The same as in Fig.~\ref{v_dotm_M}, but using HCT models and a sticking coefficient of silicates of 0.2.
}
\label{v_dotm_M03}
\end{figure}

\begin{figure*}
\centering
\includegraphics[width=0.85\textwidth,angle=90]{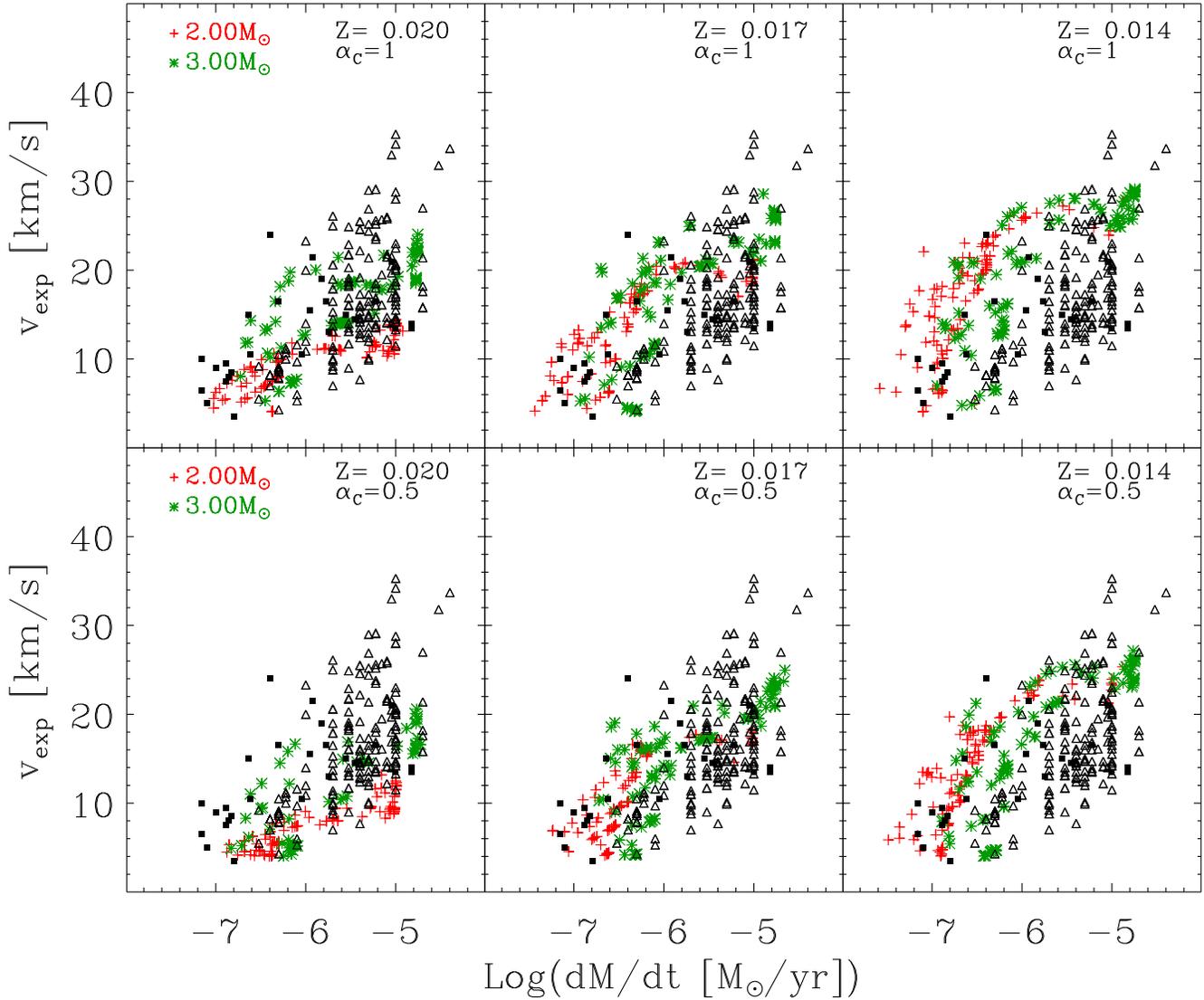}
\caption{Expansion velocities of circumstellar outflows against mass loss rates
of variable C-stars. Observations of Galactic C-stars by \citet{Loup93} (black triangles) and \citet{Schoier13} (black squares)
are compared with predicted expansion velocities for a few selected TP-AGB tracks of different
initial metallicity,  $Z=0.02, Z=0.017$ and $Z=0.014$.
The adopted sticking coefficient of amorphus carbon dust is specified in each panel.}
\label{v_dotm_C_Z}
\end{figure*}

In Fig.~\ref{v_dotm_M}, as well as in Figs.~\ref{v_dotm_large} to \ref{v_dotm_C_Z}
where we compare predictions with
observations, the evolutionary tracks for TP-AGB stars with various masses
are represented with a discrete number of points, selected from
a  randomly generated uniform distribution of  ages that
samples the entire TP-AGB phase of each star.
The number of points is not set equal to the total number of observed objects
because our aim is only to highlight
the regions where TP-AGB  stars are expected to spend most of their evolutionary time,
and not that of performing a population synthesis analysis.
This latter would require the convolution with the initial
mass function and the star formation rate, which is beyond the goal of this paper.

In the left panel of Fig.~\ref{v_dotm_M} we compare the data with the LCT models
(with chemisputtering) while, in the right panel,
we consider the HCT models (with only sublimation).
It is immediately evident that the models with
efficient chemisputtering cannot reproduce the observed velocities.
The predicted velocities never exceed 10 km~s$^{-1}$.
Moreover, after an initial rise, they saturate or even decrease
at increasing mass loss rate, failing to reproduce the observed trend.
This long standing discrepancy has challenged many theoretical investigations
\citep[e.g.]{Ireland06, Woitke06}.
As discussed also by FG06 this problem is largely insensitive to the adopted opacities.
Even using different opacity data, \citep{Jones76, Ossenkopf92,  Dorschner95} in the revised version of \citet{LeSid96},
we cannot reproduce the observed relation.

Recently a solution to this discrepancy has been advanced by  \citet{Hofner_size08}
who was able to obtain the needed acceleration by combining (i)  photon scattering from
large iron-free silicate grains with (ii) a high
sticking coefficient ($\alpha_{\rm sil}=1$).

An alternative  explanation of the discrepancy between observed and predicted
terminal velocities is naturally provided by our HCT models,
as can be seen in the right panel of Fig.~\ref{v_dotm_M}.
With a higher condensation temperature for silicates,
dust formation can take place in inner regions of the CSE,
where the acceleration term is larger.
The observed trend of velocity with mass loss rate is now satisfactorily reproduced,
with the predicted values only slightly lower than observations.
In order to improve the comparison, other important input parameters could be varied such as,
the average dust size and the sticking coefficients.

The need for using large iron-free grains is thoroughly discussed by \citet{Bladh12}.
In order to investigate the effects of large grains in our HCT models
we compute a few models decreasing the number of seeds to $\epsilon_s$=10$^{-14}$.
In these runs we use an opacity suitable for iron-free silicates
(forsterite and enstatite) taken from
\citet{Jager03}.
With a lower number of seeds we obtain grain sizes larger than $\sim$0.3 $\mu$m.
The models are plotted in Fig.~\ref{v_dotm_large}.
The comparison with the observed velocities is clearly improved, with respect to the results
shown in the right panel Fig.~\ref{v_dotm_M}.
The models show two separate branches.
The lower one is populated by stars with a current mass not
yet highly affected by mass loss. The upper sequence is populated by models
in the latest stages of the evolution, when the current mass is
significantly lower than the initial one.
This sequence is present in all our computed TP-AGB tracks but
it is more evident in the case of large grains.
Though the agreement with the data is not yet perfect,
inspection of Fig.~\ref{v_dotm_large} shows evidence that this effect could be real.
Indeed there is a sequence of stars that runs above and detached from the bulk of the
data. These data could represent stars in the latest stage of the TP-AGB phase.

The effects of changing only the sticking coefficient are shown in Fig.~\ref{v_dotm_M03}.
Here we use HCT models with standard sizes and opacity but with a modest variation of the sticking coefficient,
from 0.1 to 0.2. This variation is reasonable given that the sticking coefficients for
circumstellar conditions are not experimentally well constrained \citep{Nagahara09}.
The agreement with the data is also good, indicating that
with a larger condensation temperature it may not be necessary to
invoke iron-free grains.
\subsubsection{C-stars}
The comparison with C-stars is shown in Fig.~\ref{v_dotm_C_Z}.
Observed terminal velocities and mass loss rates of Galactic C-giants
are taken from \citet{Loup93} (black triangles) and \citet{Schoier13} (filled black squares).
The observed velocities range from a few km~s$^{-1}$ to about 35 km~s$^{-1}$ for mass loss rates between
$\simeq$10$^{-7}$~M$_\odot$yr$^{-1}$ and $\simeq$10$^{-5}$~M$_\odot$yr$^{-1}$.
LCT models of C-stars with initial mass between 2~M$_\odot$ and 3~M$_\odot$, metallicity $Z=0.02$ and sticking coefficient
of amorphous carbon $\alpha_{\rm C}=1$ (top left panel) reproduce fairly well the
observed diagram.  This result is already known
because the opacity of amorphous carbon is by far larger than that of silicate dust.

The other two upper panels in Fig.~\ref{v_dotm_C_Z} show the effects of lowering the initial metallicity from $Z=0.02$ to $Z=0.017$ and to $Z=0.014$.
At a given mass loss  rate, the predicted terminal velocities increase at decreasing Z
and while for $Z=0.02$ the models perform fairly well, at lower metallicity there is a tendency to run
above the region occupied by the bulk of the data.

In particular,  for $Z=0.014$
many models fall in a region with detectable mass loss rates
but with velocities significantly higher than observed.
To explain this effect we notice that, as shown in Figs.~\ref{fig_agbmodz001}-\ref{fig_agbmodz02},
at decreasing metallicity C-stars of the same mass not only reach a higher C/O ratio, but also a larger
carbon excess, $\epsilon_{\rm C}$-$\epsilon_{\rm O}$.
Indeed, for the metallicities considered in Fig.~\ref{v_dotm_C_Z},
the maximum C/O attained in the models of 2~M$_\odot$ and 3~M$_\odot$ are
C/O$=1.24$ and 1.46 for $Z=0.02$, C/O$=1.52$ and 1.70 for $Z=0.017$,
C/O$=1.95$ and 2.06 for Z$=0.014$, while the maximum carbon excess is
1.92$\times$$10^{-4}$ and 3.52$\times$$10^{-4}$ for $Z=0.02$,
3.63$\times$$10^{-4}$ and 4.57$\times$$10^{-4}$ for $Z=0.017$
and 5.39$\times$$10^{-4}$ and  5.58$\times$$10^{-4}$ for $Z=0.014$, respectively.
With a larger carbon excess, the amount of amorphous carbon
that can be produced in our scheme is also larger and the wind experiences a higher acceleration.
To see how this result depends on the adopted value of the sticking coefficient of the amorphous
carbon we have recomputed the above dust evolution sequences
with both $\alpha_{\rm C}=0.5$ and $\alpha_{\rm C}=0.1$.
With the latter value we could not obtain C/O ratios within the range
that can produce realistic expansion velocities \citep{Mattsson10}.
With the intermediate value $\alpha_{\rm C}=0.5$ one can still get
a good agreement with the data, as shown by the lower panels of Fig.~\ref{v_dotm_C_Z}.
With $\alpha_{\rm C}=0.5$ the best agreement between data and simulations
is achieved for $Z=0.017$ and, as in the case with sticking coefficient $\alpha_{\rm C}=1$, $Z=0.014$
seems to be a too low metallicity for the Galactic C-stars.
Clearly, there is a degeneracy between the carbon excess and the sticking coefficient,
so that we expect that even for $Z=0.014$ there will be a value
of the sticking coefficient, $0.1<\alpha_{\rm C}<0.5$,
that is able to reproduce the observed data.
Nevertheless, given that a metallicity between $Z=0.02$ and $Z=0.017$ is fairly representative
of the possible values for C-stars of the Galactic disk
and in  absence of more accurate experimental determinations,
we consider $0.5\leq\alpha_{\rm C}\leq1$ the most plausible range for
the sticking coefficient of amorphous carbon.

\begin{figure}
\includegraphics[width=0.4\textwidth]{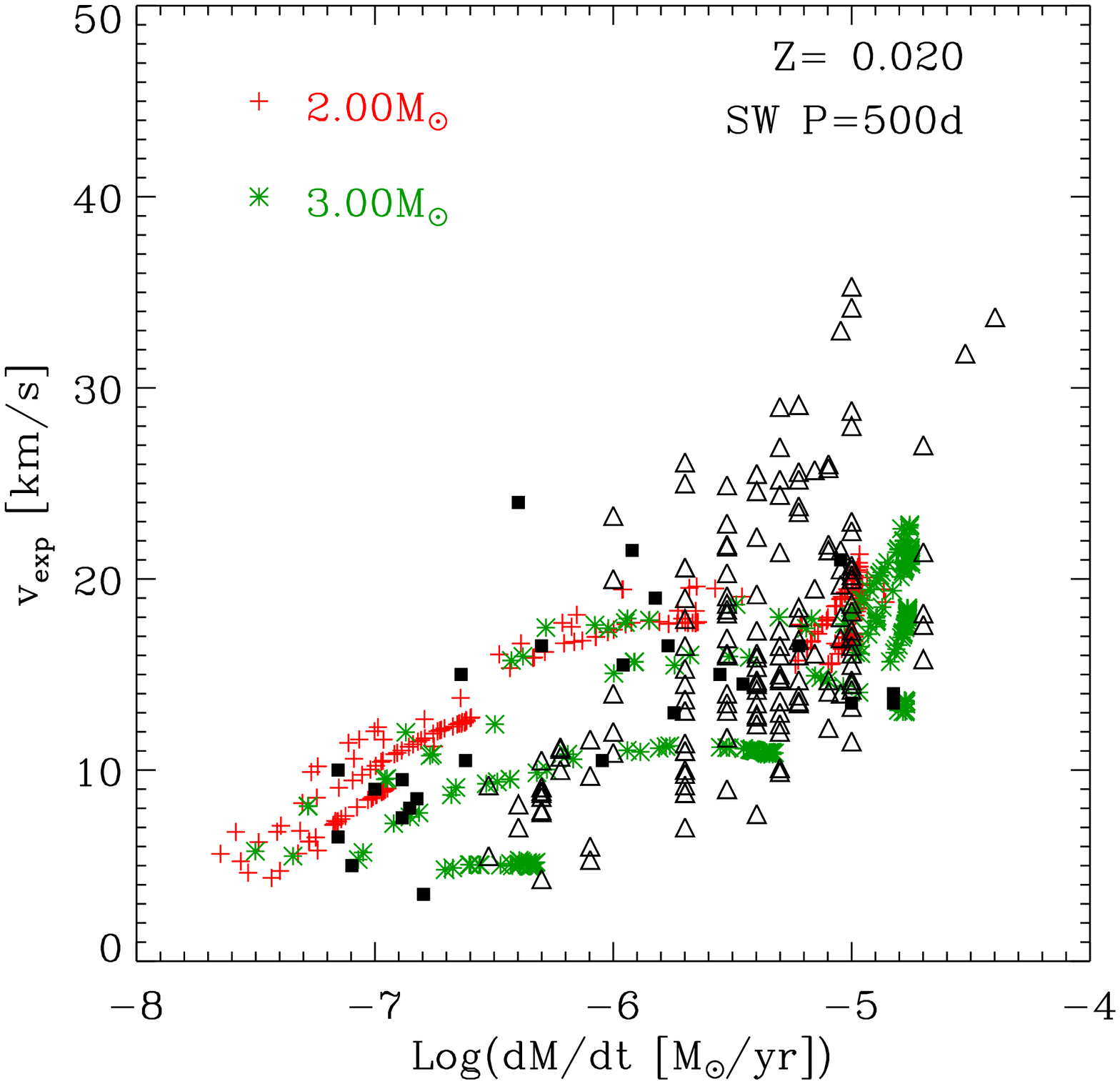}\\
\includegraphics[width=0.4\textwidth]{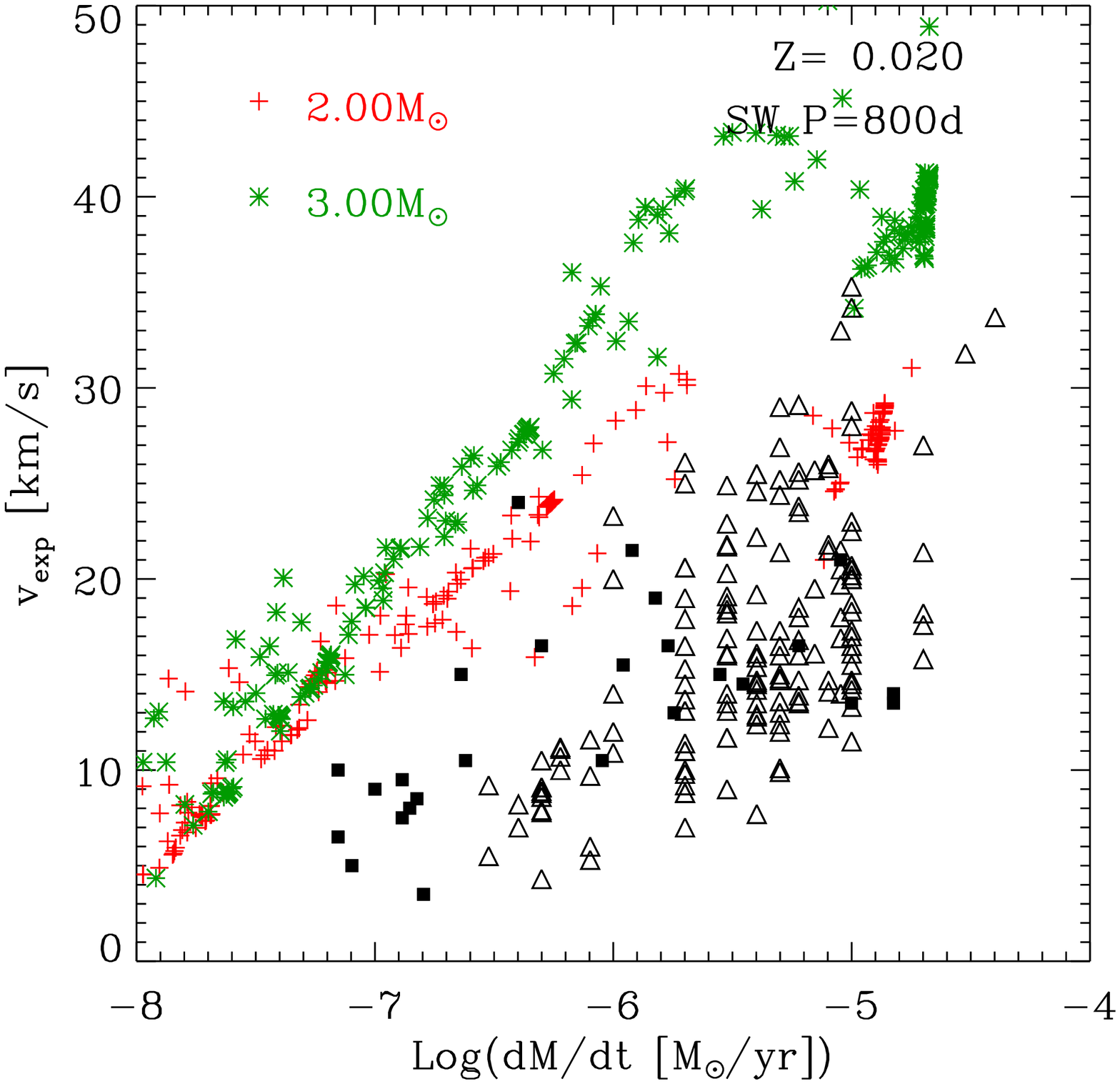}\\
        \caption{Impact of different laws of mass loss rate on the expansion velocities of CSEs of C-stars.
 Observations of Galactic C-stars by \citet{Loup93} (black triangles) and \citet{Schoier13} (black squares)
are compared
with those predicted by models of $Z=0.02$ for two other different mass loss rate recipes. \textit{Upper panel}:
the \citet{Vassiliadis93} mass loss law in its original formulation. \textit{Lower panel}: the \citet{Vassiliadis93}
mass loss law but delaying the onset of the super-wind to a pulsation period P = 800 days \citep{Kamath11}.}
\label{v_dotm_C}
\end{figure}

It is also interesting to see the effects of adopting
different mass loss prescriptions, but maintaining the same prescription for the efficiency of the third dredge-up \citep{Karakas_etal02}.
In the upper panel of Fig.~\ref{v_dotm_C} we show the effects of
adopting the mass loss rate prescription
of \citet{Vassiliadis93} in its original formulation, with a superwind phase starting at a
pulsation period
$P= 500$ days.
The general agreement is satisfactory, similar to the results in Fig.~7 (top panel).

Instead, delaying the onset of the superwind to P=800 days as suggested by \citet{Kamath11},
we predict expansion velocities that are too high compared to the bulk of the observed data,
for $10^{-6}<\dot{M}<10^{-5}$.
In this case, TP-AGB models suffer more third dredge-up events, which lead to larger
C/O ratios, i.e. 1.77 and 2.47 for the 2~M$_\odot$ and 3~M$_\odot$ models, respectively.
Lower velocities may be obtained with models of higher Z, but this would conflict with the observed oxygen
abundances of the Galactic C-stars which are slightly sub-solar.
Within our adopted scheme for C-dust formation, a delayed super-wind can still produce a reasonable agreement with observations,
but invoking a lower efficiency of the third dredge-up at the same time.

It is also worth mentioning that there is observational evidence that
C-stars in the Small Magellanic Cloud (SMC),
despite having very similar molecular C$_2$H$_2$ and HCN NIR band strengths compared to
C-stars in the Large Magellanic Cloud (LMC), show a lower intrinsic dust attenuation \citep{vanloon08}.
This would confirm earlier suggestions that also for C-stars the dust-to-gas ratio
should decrease with the initial metallicity, and this would directly affect the predicted velocities.
In particular \citet{vanloon00}, combining
scaling laws provided by spherically symmetric stationary dusty winds with other observational evidence,
derived a linear relation between the dust-to-gas ratio and the initial metallicity of C-stars.
While in M-giants a correlation with metallicity is naturally explained by the secondary nature of the
dust key-elements such as silicon and magnesium,
for C-stars this would mean that the efficiency of converting C-molecules into C-grains decreases
with the initial metallicity \citep{vanloon08}.  This would point against
the homogeneous growth assumed in this paper, and would be more in favour of
a heterogeneous growth on metal carbides (TiC, ZrC, MoC),
as already discussed  in section \ref{cstarmodels}.

In summary,  we can draw the following conclusions: expansion velocities
of C-stars depend on the interplay between mass loss and third dredge-up,
and are affected by the uncertainties in the sticking coefficients and the details
of underlying nucleation process.
As a result, some degree of degeneracy affects the predictions.
At the same time, the observed velocities may offer a powerful tool
to calibrate the above processes, provided that
other independent observational constraints
are considered jointly  (e.g. measurements of C/O ratios,
effective temperatures, mass loss rates, lifetimes from star counts, etc.).\\

\subsection{Condensation fractions, composition, dust sizes and dust mass loss rates}

\begin{figure*}
\centering
\includegraphics[angle=90,width=0.95\textwidth]{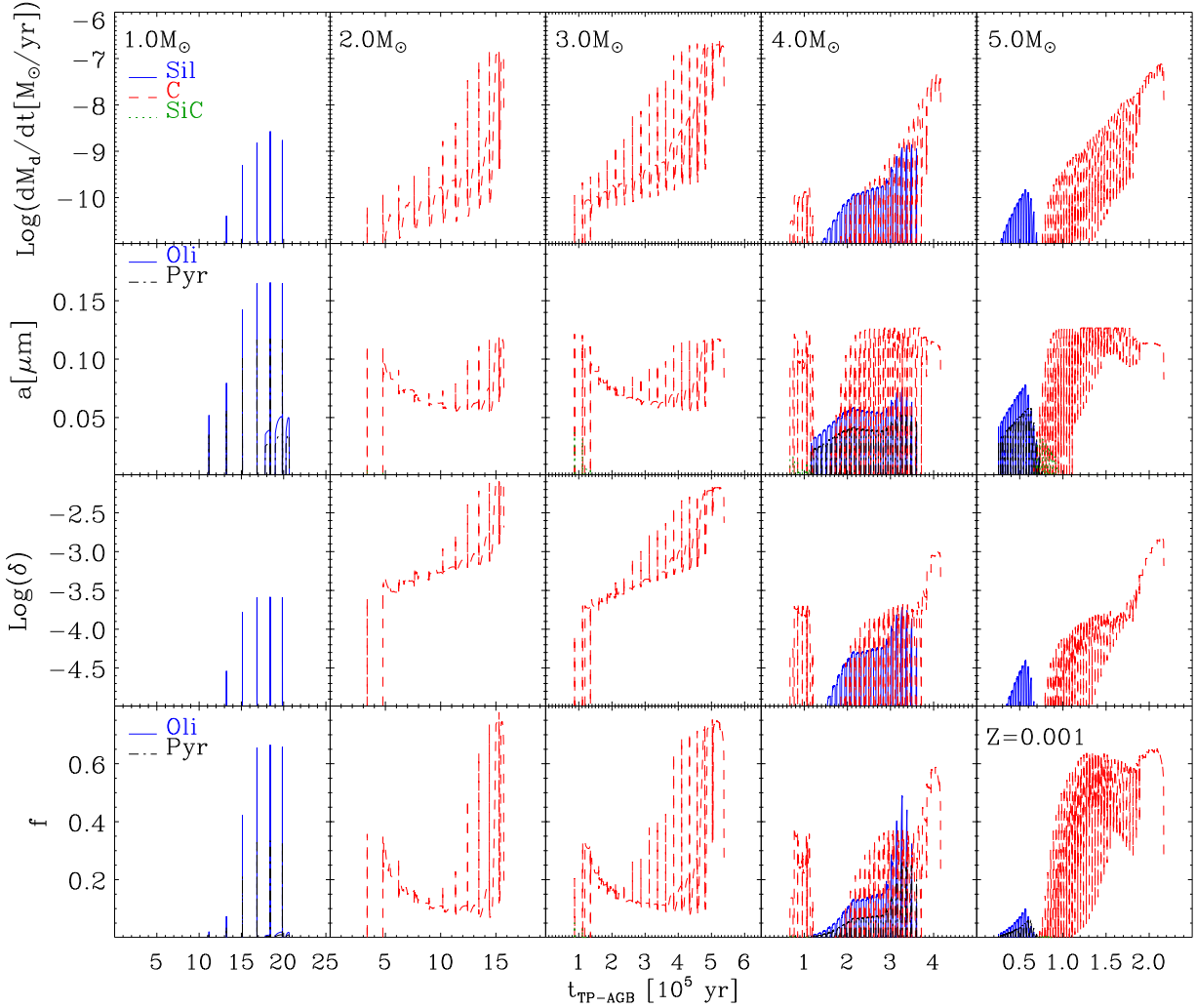}
\caption{Dust properties of selected models of initial metallicity $Z=0.001$,
for various initial masses, as shown in the upper panels.
From top to bottom we depict the dust mass loss rates in M$_\odot$yr$^{-1}$,
the dust sizes in $\mu$m, the dust-to-gas ratios $\delta$,
and the dust condensation fractions $f$, respectively.
The main dust species are silicates (blue lines), amorphous carbon (red lines) and
SiC (green lines).
In some panels silicates are separated into olivine type dust (blue lines)
and pyroxene type dust (black lines) as indicated in the insets. }
\label{fig_agbmod_dustz001}
\end{figure*}

\begin{figure*}
\centering
\includegraphics[angle=90,width=0.95\textwidth]{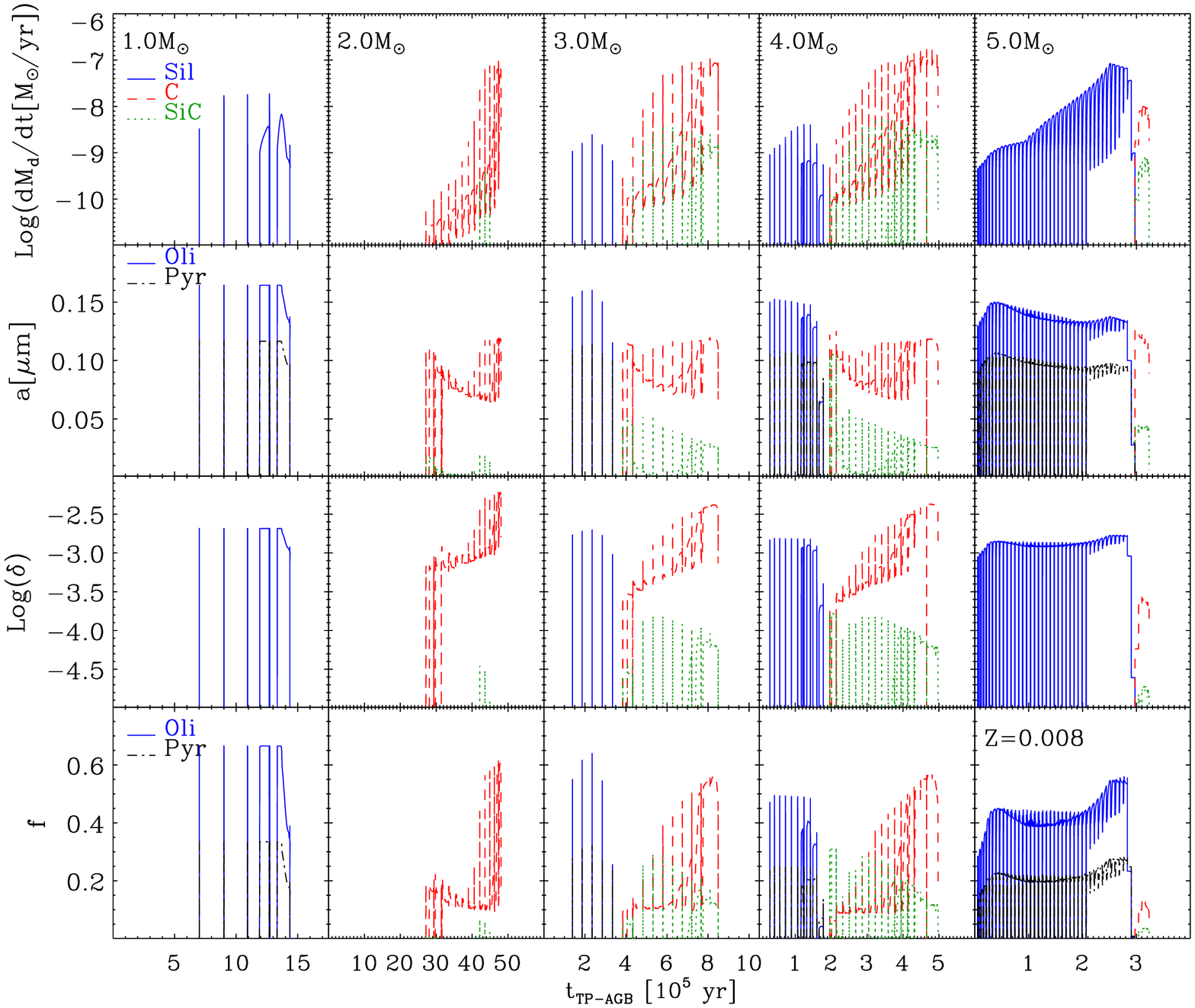}
\caption{The same as in Fig.~\ref{fig_agbmod_dustz001},
but for initial metallicity $Z=0.008$.}
\label{fig_agbmod_dustz008}
\end{figure*}

\begin{figure*}
\centering
\includegraphics[angle=90,width=0.95\textwidth]{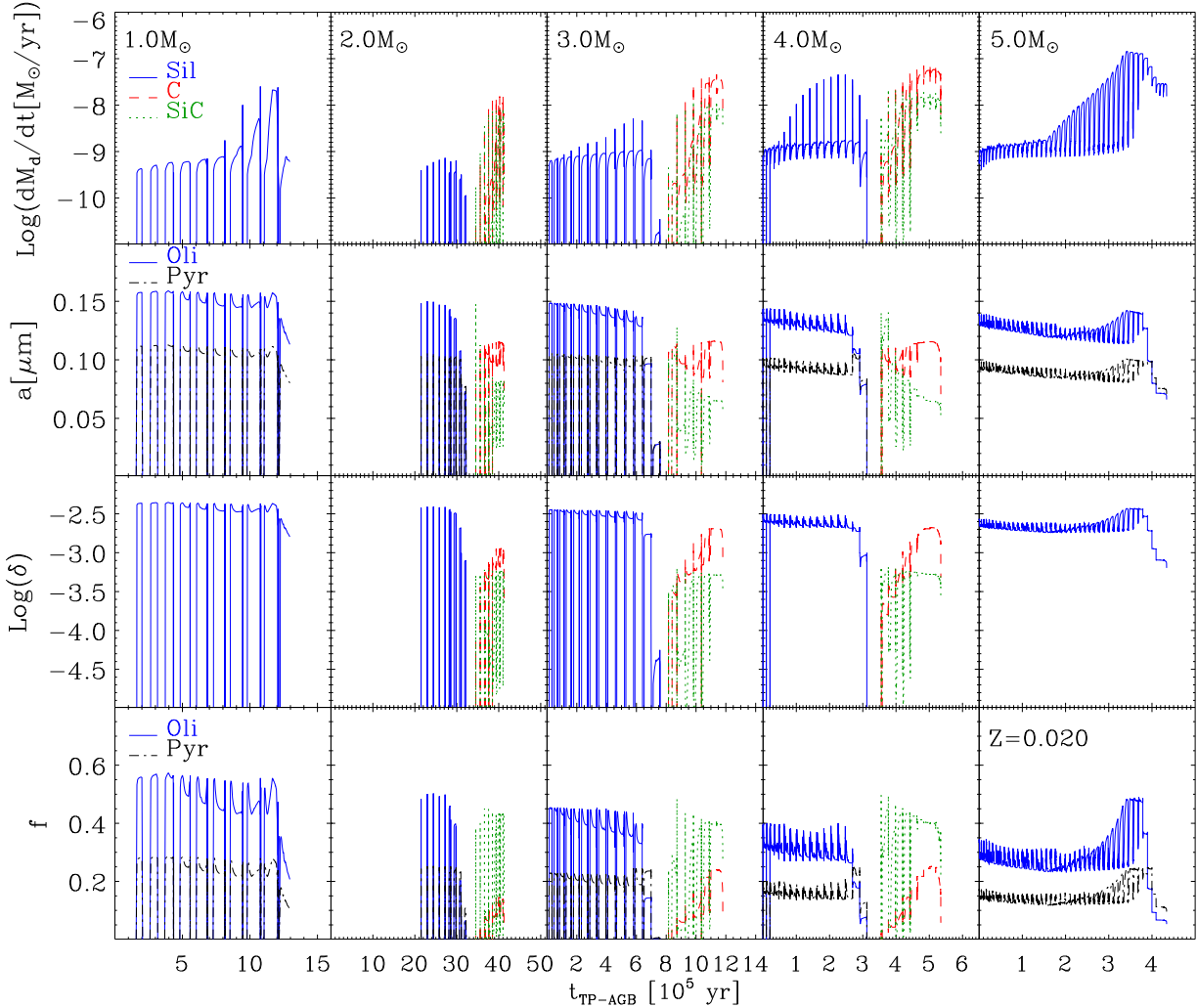}
\caption{The same as in Fig.~\ref{fig_agbmod_dustz001},
but for initial metallicity $Z=0.02$.}
\label{fig_agbmod_dustz02}
\end{figure*}

Figures \ref{fig_agbmod_dustz001}, \ref{fig_agbmod_dustz008} and \ref{fig_agbmod_dustz02} show the evolution of
the dust mass loss rates, the dust sizes, the dust-to-gas mass ratios and the dust condensation fractions
for the three different metallicities considered in this paper and for several initial masses.
We show only the results of the HCT models computed with $\alpha_{\rm sil}=0.2$, that provide the best agreement with observations in
terms of expansion velocities for M-stars.
For sake of simplicity, the mass loss rates and the dust-to-gas ratios are shown only for silicates, amorphous carbon and SiC.
As for grain sizes and condensation fractions, we distinguish between pyroxene and olivine.
Referring back to Figs.~\ref{fig_agbmodz001}-\ref{fig_agbmodz02}, it is clear that all these quantities are
modulated by the thermal pulse cycles, over which  significant
changes in $L_*$, $\Teff$, and $\dot{M}$ are expected to take place.
The chemical type of the main dust species (either silicates of carbonaceous dust)
is essentially controlled by the C/O ratio, hence it depends on
the interplay between third dredge-up which tends to increase the
carbon excess, and HBB that converts carbon, and even oxygen at lower metallicities, into nitrogen.
Stars with mass M$\leq1.5-2$~M$_{\odot}$, corresponding to item a) in section
\ref{tpagbmodels}, will produce mainly olivine and pyroxene.
At the lower metallicity the condensation fractions are very low and
these stars are able to give rise to a dust driven wind
only during the peak luminosity associated with the thermal pulses.
At increasing metallicity,  the condensation fractions become
larger and the stars are able to produce a dust driven wind also during the inter-pulse phase.
The dust composition of the stars with masses of $\approx 2-3$~M$_{\odot}$, whose evolution has been described in
item b) in section \ref{tpagbmodels}, is initially that of an M-giant,
i.e. mainly composed of silicates.
During this phase the C/O ratio increases because of the third dredge-up, while remaining below unity.
Therefore, the excess of oxygen ($\epsilon_{\rm O}-\epsilon_{\rm C}$),
available to be locked into silicates progressively decreases.  However, since the key-element is silicon,
the dust-to-gas ratio is not affected by this variation until eventually C/O$\sim$1. At this stage there is a
visible drop in silicate production.
Once the C/O ratio becomes larger than $1$, carbonaceous dust is produced, initially in the form of SiC.
In fact, SiC condenses at a higher gas temperature than amorphous carbon, but
its abundance is limited either by the silicon abundance or by the carbon excess if
$\epsilon_{\rm C}$-$\epsilon_{\rm O}\leq\epsilon_{\rm Si}$. When this latter condition is fulfilled,
almost all the carbon available condenses into SiC,
as it can be seen at the first thermal pulse of the C-star phase of the 2~M$_\odot$ and 3~M$_\odot$ models at solar metallicity.
As soon as $\epsilon_{\rm C}-\epsilon_{\rm O} > \epsilon_{\rm Si}$, amorphous carbon becomes the
dominant species as the abundance of the SiC is now controlled by the abundance of silicon.
We note that the mass loss rate during the C-star phase is about two orders of magnitude larger
than in the previous M-star phase, so that the integrated dust ejecta are mainly in the form of amorphous carbon.
The dust evolution of the more massive stars at low and intermediate metallicity
is similar to the previous case but, because of the very efficient HBB there is a
rapid initial decrease of the C/O ratio. Since the key-element
of silicate dust is silicon, the gas-to-dust ratio is unaffected by this variation.
The stars evolve at higher luminosities and cooler effective temperatures and the
mass loss rates, in both gas and dust, increase significantly.
In the case of intermediate metallicity the star  becomes carbon-rich in its last evolutionary stages
when HBB is extinguished. Since at this stage only few more third dredge-up episodes
may still take place, the contribution of carbonaceous dust to the
integrated ejecta is negligible compared to that of silicates.
In the case of the model of $M=5$~M$_{\odot}$ and $Z<0.001$
that may be considered representative of the more massive TP-AGB stars of low metallicity,
corresponding to item c) in section \ref{tpagbmodels},
the HBB is so efficient that
also the ON cycle is active, causing the partial conversion of oxygen into nitrogen.
Thus, after an initial drop of both C and O, the C/O ratio
quickly becomes larger than unity. As a consequence, silicate dust production is negligible
and the main product is amorphous carbon. SiC dust is also minor in this model.
Finally the model with $M=4$~M$_{\odot}$ and $Z=0.001$
presents several characteristics at a time. It begins as an M-giant but
its silicate condensation fraction is not high enough to power a dust driven wind.
It then becomes a carbon star, item b), but, after a while, the HBB
is enough to convert it back again into an M-giant, this time with a dust driven wind,
item d) in section \ref{tpagbmodels}. After this stage
its dust evolution is characterized
by the quasi-periodic transitions across C/O$=1$ from both directions,
caused by the alternating effects of
the third dredge-up (C/O $\uparrow$) and HBB (C/O $\downarrow$), corresponding to item e).
Accordingly, the main characteristics of this star are the concomitant
presence of both silicate and carbonaceous dust in the same thermal pulse cycle, because
the two crossings are experienced within the same cycle.
During the C-star phase the main product is amorphous carbon even if C/O$\sim$1,
because of the very low silicon abundance.

In summary, on the basis of our results,
the emerging picture concerning dust production in single stars is as follows.
At $Z=0.001$ silicates dominate dust production only at the lowest masses (M$\sim$1~M$_{\odot}$)
where the third dredge-up process is absent. However the corresponding dust mass loss rates are very low.
At larger masses, amorphous carbon is
the main dust species, though its amount may change depending on the relative
efficiencies of the third dredge-up and HBB.
At increasing metallicity, the third dredge-up and HBB combine in such a way that
the mass range for carbon dust production becomes narrower.
At $Z=0.02$, carbon dust is produced between 2~M$_{\odot}$ and 4~M$_{\odot}$
whereas, in all other masses, silicate dust is the main product.
This result is almost independent from whether chemisputtering is included or not but,
obviously,  in the case without chemisputtering, silicates tend to condense  more
efficiently.
The differences between LCT and HCT models for silicates are more pronounced
at mass loss rates below 10$^{-6}$M$_\odot$yr$^{-1}$
while, at higher mass loss rates,
the models reach approximately the same condensation degree.
These are also the phases
that dominate the total dust production.

The sizes $a$ of dust grains span a wide range of values during the TP-AGB evolution,
from a few hundredths of a micron up to a maximum, typically around 0.15 $\mu$m, for all stellar
masses and metallicities, as shown in Figs.~\ref{fig_agbmod_dustz001}-\ref{fig_agbmod_dustz02}.
The broad range of $a$ spanned during the TP-AGB phase depends mainly on the efficiency
of mass loss, as this latter determines the volume density of the
key-elements which, in turn, regulate the growth rates
(see Eq.~\ref{growth}).
Interestingly, the maximum value of the size is almost independent from the metallicity of the star.
This is due to the fact that both the initial number of seeds (Eq.s \ref{nseeds_M} and \ref{nseeds_C})
and the total amount of dust that may condense, scale linearly
with the metallicity  or with the carbon excess.
Indeed, by decreasing the number of seeds to $\epsilon_s$=10$^{-14}$
in a few test models, we obtain a maximum size $\geq$0.3~$\mu$m, as expected from the above
simple scaling.

\begin{figure*}
\centering
\includegraphics[angle=90, width=0.75\textwidth]{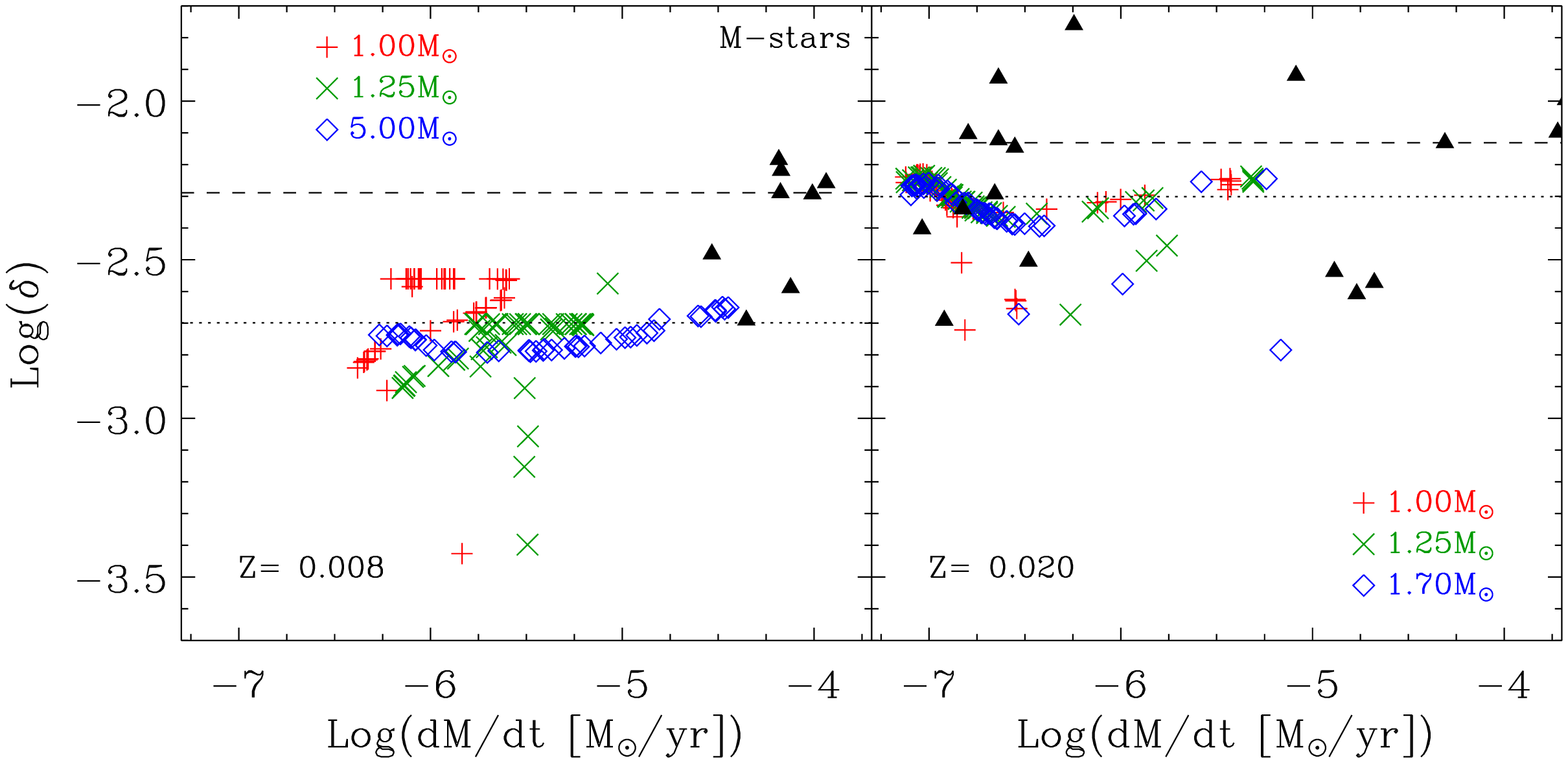}
\caption{Observed and predicted dust-to-gas ratios of Galactic and LMC M-stars. \textit{Left panel}: LMC data \citet{Marshall04}
are compared with models of $Z=0.008$ and initial masses of 1, 1.25 and 5 M$_\odot$.
The dashed line indicates the median value of the data ($1/194$), whereas the dotted line
represents the usually assumed value of the dust-to-gas ratio $1/500$ \citep{vanloon05}.
\textit{Right panel}: comparison
between models of $Z=0.02$ and initial masses of 1, 1.25 and 1.7 M$_\odot$ with Galactic data from \citet{Knapp85}.
The dashed line indicates the median value of the data ($1/135$), whereas the dotted line
represents the usually assumed value of the dust-to-gas ratio $1/200$ \citep{Groenewegen_dg98}. See text for details.}
\label{gal_dg_mstars}
\end{figure*}
The lower panels in Figs.~\ref{fig_agbmod_dustz001}-\ref{fig_agbmod_dustz02}
show
the dust-to-gas ratios $\delta$, and the condensation fractions $f$, respectively.
The dust-to-gas ratios of CSE models can be compared
with observations when
independent data exist for both the dust mass loss and the gas mass loss rates.
Unfortunately,  the dust-to-gas ratio is difficult to measure mainly due to the
difficulties of obtaining accurate measurements of the gas mass loss rates.
For example, for the extragalactic C-stars, for which CO observations are challenging,
the dust-gas-ratio is almost always assumed and used to derive the total mass loss rate
from mid infrared observations.
An important issue is whether and how this ratio depends on metallicity and
chemistry. There are indications that for M-giants it decreases  almost linearly
with the metallicity \citep{Marshall04}, while for the C-stars the
situation is more intriguing because carbon is enhanced by the  third dredge-up,  and so
dust production may not reflect the initial metallicity as in M- giants.
If also carbon formed from heterogenous nucleation on metal carbides, as suggested by
some recent observations \citep{vanloon08}, then a behavior similar to that
shown by M-giants should be expected. If a significant fraction of carbon
formed from homogeneous growth, as suggested by the absence of metallic seeds
in the nuclei of several meteoritic graphite spherules \citep{Bernatowicz96},
then the run of the dust-to-gas ratio may be more dependent
on the evolution along the AGB.
Typical values  assumed for the CSE dust-to-gas ratios are 1/200 for the Galaxy,
1/500 for the LMC and 1/1000 for the SMC \citep{Groenewegen98, vanloon05}.
 Dust-to-gas ratios  twice the quoted values are also quite common
\citep{Groenewegen09, Woods12}.

A noticeable effect that can be already seen in
Figs.~\ref{fig_agbmod_dustz001}-\ref{fig_agbmod_dustz02}
is that in M-giants the dust-to-gas ratio ($\delta$) shows only a mild dependence
on the evolutionary status of the star. Of course, there are strong variations
near the peak luminosity after each thermal pulse but,
during the inter-pulse phase, its value runs almost flat.
In contrast, in the case of C-stars, $\delta$ generally increases during the evolution both at the
peak luminosity and at each inter-pulse cycle. This mainly reflects our choice of
adopting a homogeneous growth process for carbon.
The comparison with Galactic M-giants (right panel) and with LMC data (left panel) is shown in Fig.~\ref{gal_dg_mstars}.
For Galactic M-stars we use the sample by \citet{Knapp85} providing both dust and gas mass loss rates.
For LMC M-stars there exist only very few data in literature that can be used to determine $\delta$.
From OH observations of a small sample of M-giants \citep{Marshall04},
we derive the gas mass loss rates using
the prescription presented by \citet{vanderVeen89}
\begin{equation}
\dot{M}= f_{\rm OH}/f_{\rm OH\odot}\times1.8\times{10^{-7}}\sqrt{F_{\rm OH}}v_{exp}D
\label{eq_foh}
\end{equation}
where D is the distance in kpc, assumed D=50 for the LMC, F$_{\rm OH}$ is the
OH 1612-MHz maser peak intensity in Jy and v$_{exp}$ is the wind speed in km s$^{-1}$.
The latter two quantities are provided by \citet{Marshall04} while
the factor $f_{\rm OH}$, representing the conversion factor from OH
to H$_2$ abundances, is scaled with the metallicity, i.e.
$f_{\rm OH}$=~0.02/0.008$\times$ $f_{\rm OH\odot}$ ($f_{\rm OH\odot}$=1.6$\times{10^{-4}}$) in Eq.~(\ref{eq_foh}).
The dust mass loss rates are derived from the total mass loss rates
provided by \citet{vanloon05} for common stars, which are based on mid infrared spectral fitting.
We re-normalize the values to the \citet{Marshall04} velocities and multiply by the dust-to-gas ratio assumed by \citet{vanloon05}, $\delta=1/500$.
These data are compared with selected HCT models with $Z=0.02$ and $Z=0.008$ and
initial stellar masses of 1, 1.25, 1.7 and 5~M$_\odot$, the latter only for the lower metallicity.
Like in the comparison with the velocities, the models are drawn from a uniform randomly generated set of AGB ages, for each mass.
Furthermore we also take into account that the H$_2$ abundance in our wind models is $\approx$75\% of the total gas mass and, accordingly, we multiply our total mass loss rates for the same factor.
\begin{figure*}
\centering
\includegraphics[angle=90, width=0.75\textwidth]{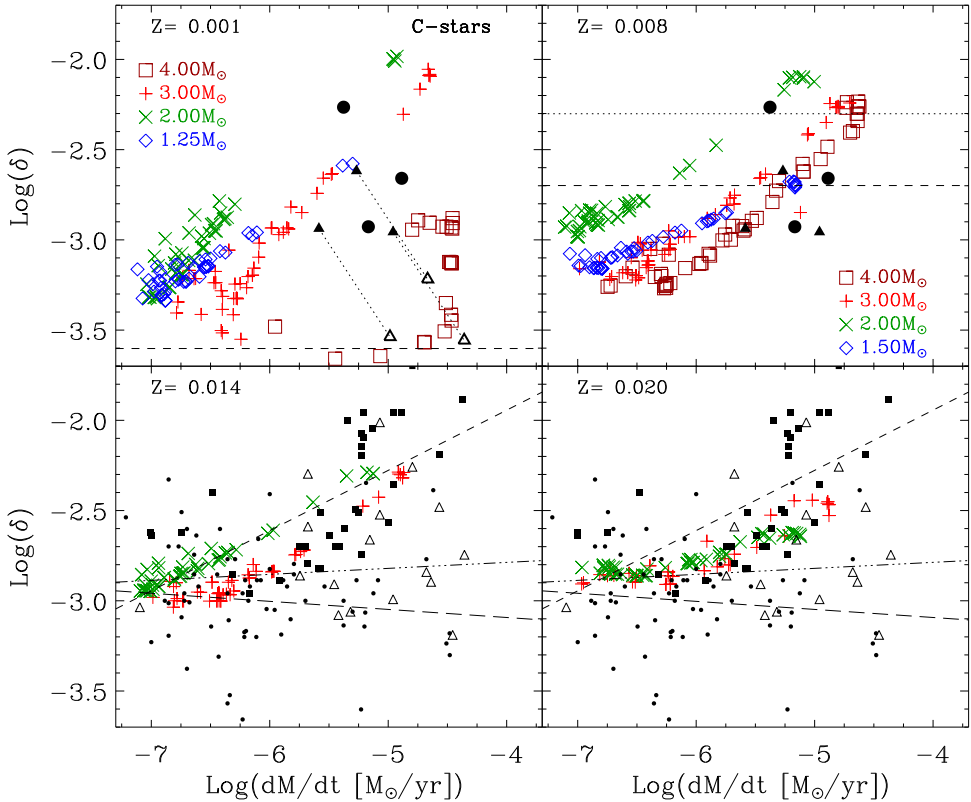}
\caption{Observed and predicted dust-to-gas ratios of C-stars.
\textit{Upper panels}: comparison
between models with $Z=0.001$ and $0.008$ and initial masses of 1.25, 1.5, 2, 3, 4~M$_\odot$ with  thick disk
C-stars (filled triangles) and Galactic Halo C-stars (filled circles) from \citet{Lagadecetal12}.
Empty thick triangles are obtained adopting an H$_2$/CO factor four times larger than that used by \citet{Lagadecetal12} for Halo C-stars.
In the left panel the dashed line represents $\delta=1/4000$. In the right panel,
the dotted line represents $\delta=1/200$ and the dashed line is
the median value of stars.
\textit{Lower panels}: comparison
between models of $Z=0.014$ and $Z=0.02$ and initial masses of 2 and 3~M$_\odot$ with Galactic data from \citet{Knapp85}
(open triangles), \citet{Groenewegen98} (filled boxes)
and \citet{Bergeat05} (small dots). In these panels the lines are best fits to the logarithmic values of $\delta$ and mass loss rates. Short-dashed line represents the
fit for \citet{Groenewegen98}, dot-dashed line is for \citet{Knapp85} and long-dashed line for \citet{Bergeat05}.
}
\label{gal_dg_cstars}
\end{figure*}
The dashed line in the right panel of Fig.~\ref{gal_dg_mstars} represents the median
value for the \citet{Knapp85} data, $\delta=1/135$. We prefer the median value of the data since, in the case of
such a range of values covering about one order of magnitude because of different physical conditions,
the mean value would be  biased toward the highest values ($\delta=1/106$). The resulting value is larger than that usually assumed, $\delta=1/200$
shown by the dotted line. Our models cluster around $\delta=1/200$ and
systematically underestimate the observed median value by $\approx$50\%.
We notice that in our models the total fraction of
silicon that is condensed into dust is already high, typically $\sim$0.7 (see also lower panels of
Figs.~\ref{fig_agbmod_dustz001}-\ref{fig_agbmod_dustz02})
and that, in order to reproduce the observed values at solar metallicity, all
the silicon should be locked into silicate dust, as already noticed by \citet{Knapp85}, unless
the adopted metallicity is too low.
We also notice that, as already anticipated, the models show
only a small dependence on the evolutionary phase along the AGB, represented here by the value of the total mass loss rate.
Concerning the LMC (left panel), we see that the median value of the data ($\delta=1/194$, dashed line)
is $\approx$40\% lower than that of Galactic M-giants. Instead our models cluster around
$\delta=1/500$ (dotted line), the value usually adopted for the LMC.
\citet{Marshall04} already noticed that, in order to reproduce the total mass loss rates
derived from infrared observations assuming $\delta=1/500$ \citep{vanloon05}, they had to
use a conversion factor for the LMC stars $f_{\rm OH}/f_{\rm OH\odot}=5$.
Had we used this higher conversion factor, the median of the data
would have been very near to $\sim 1/500$. However, unless the adopted metallicities for the Galactic and LMC stars
are grossly in error, our scaling  seems more robust
because oxygen is a very good tracer of the metallicity, even for non standard elemental partition.
We finally notice that the models do not reproduce the highest observed mass loss rates,
in both Galactic and LMC M-giants.
In the latter case, the mismatch is more evident because the observations are likely biased toward the
highest mass loss rates. Because also at this metallicity
the trend with the total mass loss rate (if any) is modest,
pushing the models toward higher mass loss rates would not solve the discrepancy with the observed
dust-to-gas ratios.

Predicted dust-to-gas ratios for C-stars are compared with the observed data in Fig.~\ref{gal_dg_cstars}.
Data for Galactic C-stars are shown in the lower two panels of Fig.~\ref{gal_dg_cstars}.
Open triangles refer to the data by \citet{Knapp85} (median value $1/693$, $\log\delta=-2.84$),
filled boxes to \citet{Groenewegen98} (median value $1/370$, $\log\delta=-2.57$),
and small dots to the revised compilation by \citet{Bergeat05} (median value $1/909$, $\log\delta=-2.96$).
C-stars LCT models with initial masses of
2 and 3~M$_{\odot}$ are shown in the lower right panel for $Z=0.02$ and in the lower left panel for
$Z=0.014$, respectively. In contrast with M-giants,
the models show a clear trend of increasing dust-to-gas ratio with mass loss rate, that becomes more evident at
decreasing metallicity. In order to judge whether this trend is real we have fitted with a least square routine
the logarithmic values of $\delta$ and mass loss rates. Only the data of \citet{Groenewegen98} (short-dashed line)
show a similar trend while in both the sample of \citet{Knapp85} (dot-dashed line) and in that of  \citet{Bergeat05} (long-dashed line) this trend is not observed.
Overall the models are in fairly good agreement with the observations of Galactic C-stars
though they cannot reproduce the highest observed ratios.
\begin{figure*}
\centering
\includegraphics[angle=90, width=0.75\textwidth]{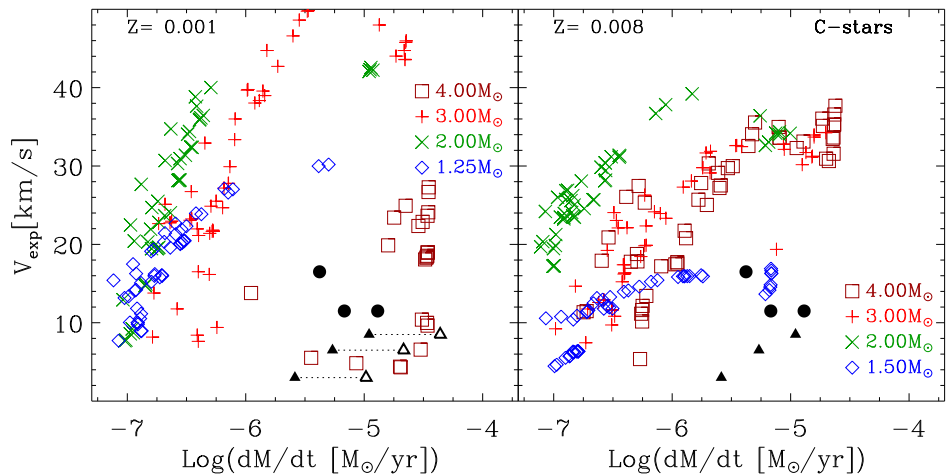}
\caption{Predicted expansion velocities of C-stars at low and
intermediate metallicities for selected
models with  $Z=0.001$ and $Z=0.008$ and initial masses of 1.25, 1.5, 2, 3, 4~M$_\odot$.
Data for thick disk stars (filled triangles) and Galactic Halo stars (filled circles), are taken from \citet{Lagadecetal12}.
At the lower metallicity  we show the effect of accounting for a H$_2$/CO conversion factor equal to four times that used by  \citet{Lagadecetal12}
for the Halo stars. See text for more details.}
\label{vexp_dotm_lowz_cstars}
\end{figure*}
The predictions of the models at intermediate and low metallicities are shown in the
upper right and left panels of Fig.~\ref{gal_dg_cstars}, respectively.
At metallicities lower than those of Galactic disk, the comparison with observations is challenged by
the lack of gas-mass loss observations in  typical subsolar environments rich in C-stars, such as the Magellanic Clouds.
We plot in Fig.~\ref{gal_dg_cstars} only the recent observations of a few metal
poor C-stars toward the Galactic Halo by \citet{Lagadecetal12}.
From radial velocities, distances
and Galactic coordinates \citet{Lagadecetal12} place their memberships in the
thick disk (filled triangles in Fig.~\ref{gal_dg_cstars}) or in the Halo (filled circles).
The nature of the Halo stars is still a matter of debate.  In our low metallicity models the minimum mass for a star
to become C-star is M$\approx$1.2M$_\odot$, while the typical turn-off mass for an old metal poor population like that of
the Halo is  M$\approx$0.9M$_\odot$. Since the lack of an efficient  third dredge-up at low masses is
a common feature of all the models of C-stars and assuming that they are correct, these stars do not belong to the Halo population
and they could have formed either on the  thick disk or outside the Galaxy
(e.g. in the Sagittarius dSph Galaxy).
Furthermore, the Halo stars show strong C$_2$H$_2$ molecular absorption bands, typical of
low metallicity stars \citep{vanloon08} and strong SiC emission, characteristic of more metal rich stars.
Since the metallicity of these stars cannot be firmly constrained, we compare their dust-to-gas ratios
with the predictions of both the intermediate and low metallicity models.
The median value of their dust-to-gas ratios, $\delta=1/456$ ($\log\delta=-2.66$), is very similar to
that assumed for intermediate metallicity stars, $\delta=1/500$, which we show as a dashed line in the upper right panel.
In the same panel the dotted line represents $\delta=1/200$, typical of solar metallicity but sometimes taken as reference also
for intermediate metallicities. For $Z=0.008$ we plot the models of initial masses 1.5, 2, 3 and 4~M$_{\odot}$.
For $Z=0.001$ the dashed line represents $\delta=1/4000$, obtained by scaling
the $Z=0.008$ usual reference value linearly with the metallicity. For $Z=0.001$ we plot the models of 1.25, 2, 3 and  4~M$_{\odot}$.
At these metallicities the trend of increasing $\delta$ with increasing mass loss rate becomes more evident.
We also notice that, at decreasing metallicity the dust-to-gas ratios at low mass loss rates
decreases while the maximum value,  reached at the highest mass loss rates, increases
with a global spread of about one order of magnitude. Thus a constant dust-to-gas ratio is not supported by our models. Furthermore
our simple random selection process indicates that at low metallicity lower values of $\delta$ are
preferred, but these values are still significantly higher than those predicted by a linear scaling with the metallicity.
We recall also that at the lowest metallicity the most massive AGB stars undergo a very efficient HBB which inhibits
the growth of the carbon excess to large values. This explains why, at $Z=0.001$, the maximum value reached by
the model of 4~M$_{\odot}$ is significantly lower than those of less massive stars.
If we adopt an intermediate metallicity for the \citet{Lagadecetal12} C-stars, their dust-to-gas ratios can be
fairly well reproduced by our models though, for most of them, there is a significant degeneracy with the initial mass.
However at this same metallicity we cannot reproduce the velocities of the Halo stars, while thick disk stars
could be compatible with the model of M=1.5~M$_{\odot}$, as shown in Fig.~\ref{vexp_dotm_lowz_cstars}.
Halo stars expansion velocities can be reproduced at the lower metallicity only by
those models in which the HBB is very efficient, such as that with M=4~M$_{\odot}$.
At this metallicity one should also correct for a  H$_2$/CO conversion factor higher than that used by  \citet{Lagadecetal12}.
This would shift the data toward higher gas mass loss rates and lower values of $\delta$
as shown by the empty thick triangles in Figs.~\ref{gal_dg_cstars} and \ref{vexp_dotm_lowz_cstars},
obtained adopting an H$_2$/CO factor four times larger than that used by \citet{Lagadecetal12}.
Notice that with this correction the Halo stars are fully compatible, in dust-to-gas ratio,
velocities and mass loss rates, with low metallicity models
where the growth of carbon excess is inhibited by an efficient HBB.
Whether this is a real effect or if it mimics a more general need of reducing the efficiency of
carbon dust formation at low metallicity, as suggested by other authors, must be analyzed by means of models that
take fully into account the heterogeneous growth of amorphous carbon on metal seeds \citep{vanloon08}.
This is the subject of a forthcoming investigation.

The dust composition of our models can be derived
from the condensation fractions plotted in the lower panel of
Figs.~\ref{fig_agbmod_dustz001}-\ref{fig_agbmod_dustz02}.
For amorphous silicates, the most abundant dust species in M-giants,
we consider the two main components, olivine  and pyroxene.
Crystalline silicates have also been detected from their MIR features
but their abundances are by far less than the corresponding amorphous phase.
Though the condensation fractions are plotted separately for olivine and pyroxene,
it is not yet possible to quantify their relative abundances observationally.

\begin{figure*}
\includegraphics[angle=90, width=0.75\textwidth]{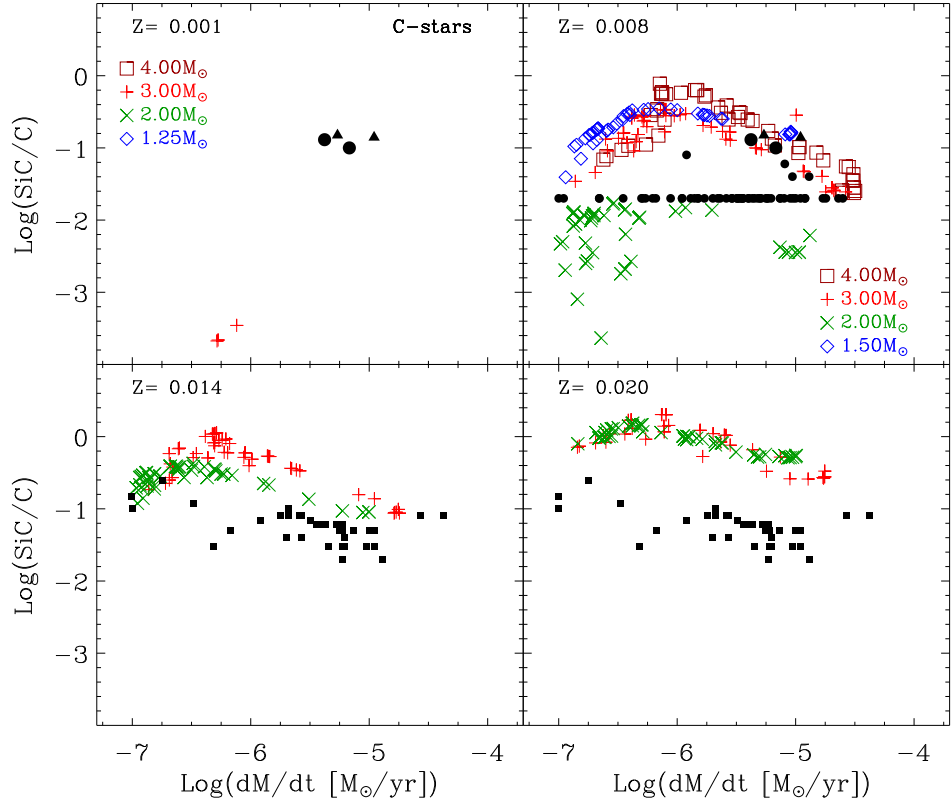}
\caption{Observed and predicted SiC/C ratios as a function of the metallicity.
\textit{Lower panels}: data for Galactic stars (filled squares), from \citet{Groenewegen98},
are compared to  models of 2 and 3~M$_\odot$ for $Z=0.02$ (right) and $Z=0.014$ (left), respectively.
\textit{Upper-right panel}:
LMC C-stars (filled small dots), from \citet{Groenewegen07},
and Halo stars (filled triangles) and  thick disk stars (large filled dots),  from \citet{Lagadecetal12},
are compared with of $Z=0.008$ and masses of  of 1.5, 2, 3 and 4~M$_\odot$.
\textit{Upper-left panel}: models of 1.25, 2, 3 and 4~M$_\odot$ and $Z=0.001$ are compared with the data from \citet{Lagadecetal12}.
}
\label{fig_sic_ac}
\end{figure*}

On the contrary, in the case of C-stars it is possible to derive the relative abundances
of the two different main components, SiC an amorphous carbon.
Fig.~\ref{fig_sic_ac} shows the comparison between existing observations of the SiC/C mass
ratio and the predictions of our models, for different metallicities.
We plot the data for Galactic stars (filled squares) from \citet{Groenewegen98}
together with our models at $Z=0.02$ in the lower right panel and at $Z=0.014$ in the lower left panel, respectively.
Data for LMC C-stars (filled small dots) from \citet{Groenewegen07} are compared with models
at $Z=0.008$ in the upper right panel. Finally, the data of \citet{Lagadecetal12}
(filled triangles for the Halo memberships and large filled dots
for the thick disk memberships)
are compared with models of low and intermediate metallicity in the upper left and upper right panels, respectively.
As can be seen from the figure our models overestimate the observed SiC/C ratios at solar metallicity.
Galactic data are mostly concentrated around 0.05-0.1 while our models may reach also SiC/C=1, as shown
by the dust-to-gas ratios in Figs.~\ref{fig_agbmod_dustz001}-\ref{fig_agbmod_dustz02}.
Though the comparison improves if we consider a slightly lower metallicity for the
Galactic data, $Z=0.014$, there is an evident over-production of SiC.
At $Z=0.008$ the agreement with the observations is only partially better.
The majority of the stars have SiC/C=0.02 but also a high uncertainty \citep{Groenewegen07}.
These data are consistent with AGB stars with initial mass M$\approx$2~M$_\odot$.
On the other hand data with values larger than 0.02
are well reproduced by all the other models. In particular
the putative thick disk members in the \citet{Lagadecetal12} sample (large filled dots)
are well reproduced by stars with initial mass M=1.5~M$_\odot$.
At the lowest metallicity considered here the predicted SiC/C ratios become negligibly small.
A decreasing SiC/C ratio with metallicity is indeed observed through
the measure of the strength of the SiC spectral features at
11.3~$\mu$m. This feature decreases going from the Galaxy to the LMC and to the SMC and other metal poor galaxies of the Local Group \citep{Sloan09}.
In this respect the location of the Halo stars in the $Z=0.001$ panel are surprising, as already noticed by \citet{Lagadecetal12}.

\begin{figure*}
\includegraphics[angle=90, width=0.75\textwidth]{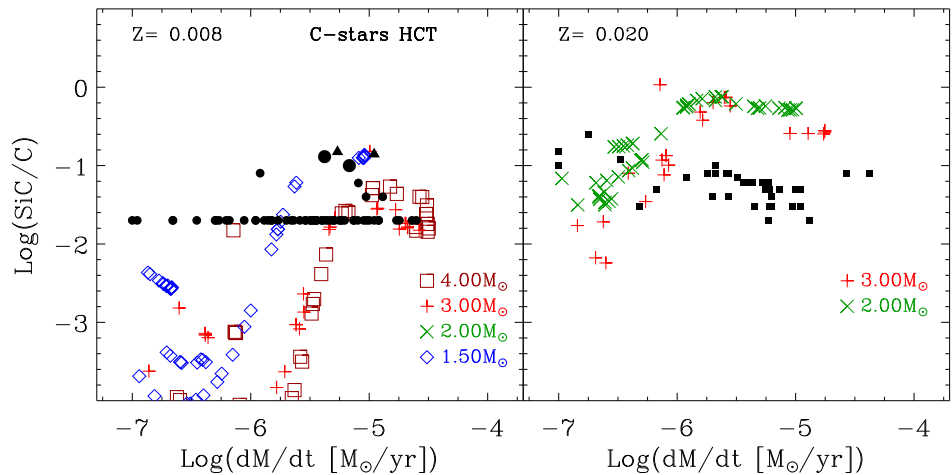}
\caption{The same as in Fig.~\ref{fig_sic_ac} for $Z=0.008$ and $Z=0.02$, but for HCT models.}
\label{fig_sic_ac_HCT}
\end{figure*}

A noticeable feature emerging from the Galactic sample is that, at increasing mass loss rates,
the SiC/C ratios decrease. This trend is shown also by the models. Inspection of our models shows that
while carbon dust-to-gas ratio, $\delta_{\rm C}$, continuously increases with the mass loss rate, $\delta_{\rm SiC}$ initially increases
and then it remains constant or even decreases. Furthermore, the maximum value reached by the models decreases with the metallicity.
While the latter effect is expected if silicon is the key element for the formation of SiC,
the global decrease with the mass loss rate is rather an effect of the internal structure and wind dynamics.
This is even more evident if we increase the temperature of carbon dust. For
the HCT models, where carbon condenses at $T_{\rm gas}$=1300~K,
the picture becomes quite different, as can be seen in Fig.~\ref{fig_sic_ac_HCT}.
At low mass loss rates the predicted SiC/C ratios are significantly lower than that of LCT models
while at high mass loss rates the value remain almost unchanged.
This behavior  is due to the fact that, at low mass loss rates, the condensation process is very sensitive to
the details of the internal structure, because the overall density is lower.
Indeed, the large opacity rise due to the condensation of carbon is always accompanied by
a strong acceleration of the wind and thus by a significant temperature and a density drop.
At lower mass loss rates the global density is lower
and an earlier condensation of carbon is enough to inhibit the formation of SiC.
On the contrary at high mass loss rates the structure is self-regulating and less dependent on the
details of the condensation process. In this case an increase of the carbon condensation temperature has an almost
negligible impact on the SiC/C ratios.

\section{Dust ejecta}
\label{sec_dust_ejecta}
By integrating the dust mass loss rate along the entire TP-AGB phase we
compute dust ejecta for the different masses and metallicities.
The ejecta refer to individual stars and they may not be representative of the
corresponding dust yields obtained after convolving with the initial mass function.
They are provided in Table~\ref{Table:ejecta} for both the LCT models
(Section~\ref{sec_chemi}), and for the HCT models (Section~\ref{sec_subli}).
The above integration is performed  irrespective of the ability of dust to drive or not the stellar wind.
In fact, we have already stressed that, for silicates and in particular in models with chemisputtering,
dust cannot always drive an outflow consistent with the adopted mass loss rate.
In these cases we must assume that there exist other possible ways to accelerate the wind,
that may finally deliver the material into the ISM \citep{Harper96}. However, these phases are characterized by
a low dust mass loss rate so that they do not contribute significantly to the total ejecta.

\begin{figure}
\centering
\includegraphics[width=0.45\textwidth]{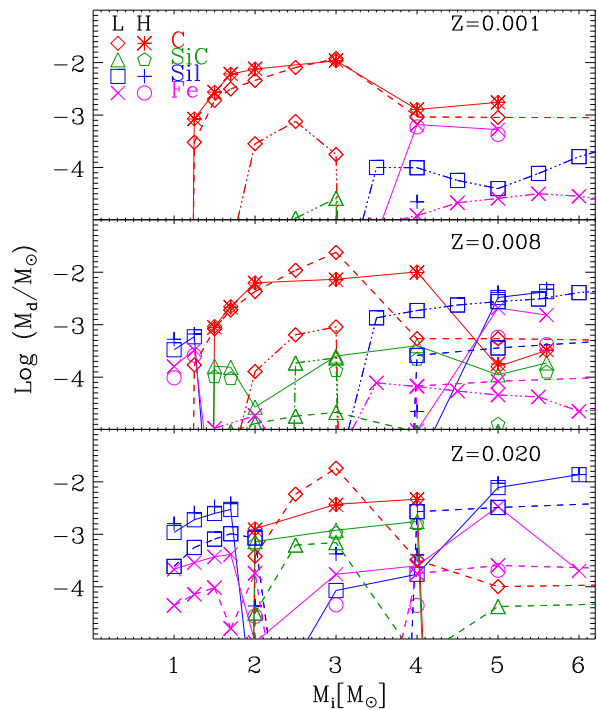}
\caption{Total dust ejecta as a function of
the initial stellar mass and for different initial metallicity.
For low-(LCT) and high-(HCT) condensation temperature  models,
we use different symbols, as indicated in the upper panel.
To facilitate the comparison with other authors, our LCT models are connected with
solid lines while, for the models with chemisputtering by FG06 we use dashed lines and,
for those of \citet{ventura12} (Z=0.001) and \citet{ventura12_2} (Z=0.008), we use dotted-dashed lines.
For our HCT models we use only the corresponding symbols.}
\label{yields}
\end{figure}

Figure~\ref{yields} shows the total dust ejecta of the main condensed compounds,
silicates, amorphous carbon, Fe and SiC, as a function of the initial stellar mass and for the three
metallicities considered here.

We first compare the results of our own models, obtained with the two different formalisms.
We remind that, in the case of C-stars, the chemisputtering process
is neglected in both schemes (see also FG06) but,
as a test case, we have recomputed the models with a higher gas condensation temperature ($=$~1300~K instead of 1100~K)
to check its effect on the total ejecta. These models are plotted as asterisks in the figure and are
referred in Table~\ref{Table:ejecta} as  HCT models. We notice also that, in the case of SiC and Fe,
possible differences in the ejecta, between LCT and HCT models, are only an indirect consequence of the
variations of the condensation temperatures of silicates and amorphous carbon, because
their dust growth schemes do not change.

As for the C-stars, the final dust ejecta mainly consist of amorphous carbon (C in Table~\ref{Table:ejecta}) and
there are not significant differences between LCT and HCT models, at any metallicity.
As far as the variation of $\alpha_{\rm C}$ is concerned, its largest effect is on the ejecta of amorphous carbon
but the maximum variation between the models
computed with $\alpha_{\rm C}=1$ and $\alpha_{\rm C}=0.5$, at solar metallicity, is only 5\%.

Differently, in the case of M-stars, we find that the silicate
ejecta of HCT models can be as much as 50~per cent  higher compared to the LCT ones.
These differences tend to vanish towards the largest mass models, since
the condensation fraction of silicates  is quite high in both classes of
 models, mostly due to the large
mass loss rates.

Thus we may conclude that, in spite of the differences in
velocity, density and temperature profiles brought about by the
adoption of either of the two formalisms,
their effect on the final dust ejecta is,  instead, rather weak.

In the same figure we plot also the results
of FG06, \citet{ventura12} and \citet{ventura12_2}, that
can be compared to our LCT models.
We remark that  FG06 TP-AGB models are based
on a compilation of analytic relations
\citep[partly taken from ][]{Marigo_etal96}  while,
\citet{ventura12} and \citet{ventura12_2} use full numerical calculations along the AGB.
The latter models also include the super-AGB
stars between 6~M$_{\odot}$ and 8~M$_{\odot}$,
that have developed an electron-degenerate O-Ne core after the C-burning phase \citep{Siess10}.

At low metallicity our results are qualitatively in good agreement
with those of FG06, since both studies predict that
the bulk of dust produced by low and intermediate mass stars is amorphous carbon.
In our models the carbon dust production
is even larger than in FG06,
by up to a factor of two in some cases.

We emphasize that the TP-AGB models computed by FG06
neglect two important aspects of stellar evolution, namely: the  break-down of
the core mass-luminosity relation due to HBB in stars with masses
$M\ga 4$~M$_{\odot}$, and the drastic changes in molecular opacities, hence in the
effective temperatures, as soon as the C/O ratio increases above unity as a
consequence of the third dredge-up. Both factors concur to a likely
overestimation of the TP-AGB lifetimes, hence to an over-exposure
 of the stellar mantle to  nucleosynthesis and mixing processes.

Therefore,  the FG06 TP-AGB models  are, by construction,
quite different from the most recent
ones of \citet{marigoetal13}. The latter models are based on
accurate numerical integrations of a complete
envelope model in which the HBB energetics and nucleosynthesis are properly
taken into account,
the initial conditions at the first TP are extracted from the
new stellar evolutionary tracks \citep{Bressanetal12},
and more importantly, they use new and accurate
equation of state and molecular opacities that
account for the continuous changes in the surface
chemical composition (see \citet{marigoetal13} for more details).

In the models by \citet{ventura12} at $Z=0.001$,
the dust production is about ten times less than
in our models.
Furthermore, while \citet{ventura12} predict that the dust ejecta are
dominated by silicates for stars with M$>3$~M$_{\odot}$, in our models
the main dust species is amorphous carbon.
This difference is likely  due to
a more efficient third dredge-up in our AGB models at this low metallicity.

At $Z=0.008$ we are again in qualitative good agreement with FG06 because
both studies predict that stars with mass below 1.5~M$_\odot$
produce silicate dust, while at larger masses the main dust
species is amorphous carbon.

However, while in FG06 models for  stars more massive than 5~M$_{\odot}$
the carbonaceous and silicate dust ejecta are comparable, in our models the ejecta are dominated by silicates.
This difference is likely due to a more efficient HBB in our models.
We also note that for $M>5$~M$_\odot$, our dust ejecta are roughly 5 times larger than
those of  FG06.

The comparison with \citet{ventura12_2} shows again that, for $1.5<M<4$~M$_\odot$, our models produce about
ten time more dust. Only for  $M\geq4.5$~M$_\odot$ their dust ejecta become comparable (and sometimes slightly higher)
to those of ours models. In \citet{ventura12_2} silicate dust production becomes efficient
above $M=3.5$~M$_\odot$ whereas, in our models this happens at about $M=4.7$~M$_\odot$.
At solar metallicity we can compare our results only with those of FG06.
The production of carbon dust is limited to the mass range  between
2~M$_\odot$ and 4~M$_\odot$, with a peak at around 3~M$_\odot$, both in our models and in those of FG06.
The larger dust ejecta predicted by FG06 are
probably due to the larger C/O ratios reached by their TP-AGB models.
In fact, FG06 do not account
for the cooling effect of the TP-AGB tracks (for C/O$>1$) , produced by the
C-rich molecular opacities, with consequent  shortening of the C-star lifetimes
\citep{Marigo_02}.

Moreover, for M$>5$~M$_\odot$,  our silicate ejecta are roughly a factor of 3 larger than FG06.
This may be related to a more efficient HBB in our models, that
lowers significantly the carbon content in the CSEs.
\begin{table*}
\begin{center}
\caption{Dust ejecta for LCT and HCT models.}
\begin{tabular}{l c c c c c c c c c c c}
\hline\hline
M$_i$ & $\log$~(age/yr) & \multicolumn{2}{c}{Sil}&  \multicolumn{2}{c}{Fe} &\multicolumn{2}{c}{Al$_2$O$_3$}&\multicolumn{2}{c}{C}&\multicolumn{2}{c}{SiC}\\
 {[M$_\odot$]} &  &\multicolumn{10}{c}{$\log$~(M/M$_\odot$)}\\
\hline
$Z=0.001$ && \scriptsize LCT &\scriptsize  HCT&\scriptsize  LCT&\scriptsize  HCT &\scriptsize  LCT &\scriptsize HCT&\scriptsize  LCT&\scriptsize  HCT&\scriptsize  LCT&\scriptsize HCT\\
\hline
1.00&  9.7 &   $-6.53$&    $-5.66$&   $-6.48$&    $-6.63$ &  $-7.44$ &  $-7.74$&   -&       -&   -         &     -\\
1.25&  9.5    & -&        -&    $-8.62$& $-8.87$&   -&      -&   $-3.07$& $-3.08$&   -    &     - \\
1.50&  9.3   & -&        -&   $-8.03$&    $-8.27$&     -&       -&  $-2.56$&    $-2.56$&  -     &     - \\
1.70&  9.1   & -&        -&  $-7.69$&  $-7.92$&     -&       -&  $-2.22$&  $-2.22$&  -     &     - \\
2.00&  8.9  & -&        -&   $-7.57$&  $-7.80$&     -&       -&  $-2.12$&  $-2.12$&  -     &     - \\
3.00&  8.5  & -&        -&   $-7.40$&    $-7.62$&     -&       -& $-1.97$& $-1.96$ & $-8.94$  & $-9.01$\\
4.00&  8.2   & $-6.92$&  $-4.65$ & $-3.18$ & $-3.23$ & $-6.03$ &  $-6.29$& $-2.89$ & $-2.89$ &   - & $-8.55$\\
5.00&  8.0 & -& $-5.75$&   $-3.28$ &   $-3.37$&     -& $-7.22$ &  $-2.76$& $-2.76$ &  -     & $-8.45$\\
\hline
$Z=0.008$&& \scriptsize LCT &\scriptsize  HCT&\scriptsize  LCT&\scriptsize  HCT &\scriptsize  LCT &\scriptsize HCT&\scriptsize  LCT&\scriptsize  HCT&\scriptsize  LCT&\scriptsize HCT\\
\hline
1.00&   9.9 &  $-3.48$& $-3.28$ & $-3.79$& $-4.01$ & $-4.53$ & $-5.24$ &   -&     -& -         &     -\\
1.25&   9.6   & $-3.23$ & $-3.18$ & $-3.48$ & $-3.58$ & $-4.28$& $-6.31$&   -&     -& -          &     -\\
1.50&   9.4  & $-7.98$ & $-7.34$ & $-6.66$& $-7.03$ & $-8.56$ & $-8.74$ & $-3.03$ & $-3.05$& $-3.79$ & $-4.00$\\
1.70&   9.3  & -&      -& $-6.90$ & $-7.64$&  -&   -& $-2.66$ & $-2.67$ & $-3.81$ & $-4.03$\\
2.00&   9.1  & -&      -& $-7.16$ & $-7.50$ & -&   -& $-2.20$& $-2.20$ &  $-4.57$ & $-6.99$\\
3.00&   8.6  & $-6.65$& $-5.95$& $-6.14$& $-6.37$& $-7.49$ & $-7.91$ & $-2.13$ & $-2.14$ & $-3.60$ & $-3.88$\\
4.00&   8.3  & $-5.69$& $-4.66$& $-5.02$&  $-5.13$ & $-6.43$ & $-6.60$ & $-1.99$& $-2.01$ & $-3.40$ & $-3.66$\\
5.00&   8.0 & $-2.48$& $-2.39$ & $-2.68$ & $-3.24$ & $-3.48$ & $-5.48$ & $-3.80$ & $-3.75$ & $-3.96$ & $-4.90$\\
5.60&   7.9 & $-2.37$& $-2.32$& $-2.82$ & $-3.38$ & $-3.39$ & $-5.73$ & $-3.49$& $-3.49$ & $-3.73$ & $-3.91$\\
\hline
$Z=0.02$&& \scriptsize LCT &\scriptsize  HCT&\scriptsize  LCT&\scriptsize  HCT &\scriptsize  LCT &\scriptsize HCT&\scriptsize  LCT&\scriptsize  HCT&\scriptsize  LCT&\scriptsize HCT\\
\hline
1.00&  10   & $-2.99$ & $-2.80$ & $-3.68$ & $-5.42$ & $-4.04$ & $-5.23$&  -&       -&   -         &     -\\
1.25&  9.7   & $-2.73$& $-2.59$ & $-3.55$&  $-5.60$& $-3.80$ & $-5.13$&  -&       -&   -         &     -\\
1.50&  9.5   & $-2.63$ & $-2.48$ & $-3.46$ & $-5.33$ & $-3.69$ & $-5.06$&  -&       -&   -         &     -\\
1.70&  9.4   & $-2.54$ & $-2.42$ & $-3.40$&  $-5.28$ & $-3.60$ & $-5.03$&  -&       -&   -        &     -\\
2.00&  9.2   & $-6.73$ & $-4.37$ & $-5.06$ & $-4.56$ & $-7.29$ & $-6.07$& $-2.89$ & $-2.92$ & $-3.14$ & $-3.20$\\
3.00&  8.7   & $-4.07$& $-3.37$& $-3.76$& $-4.34$ & $-4.75$& $-5.06$&  $-2.43$& $-2.44$& $-2.92$ &  $-2.98$\\
4.00&  8.3   & $-3.76$& $-3.40$& $-3.60$ & $-4.36$ & $-4.38$ & $-4.94$&  $-2.33$& $-2.34$& $-2.76$  &  $-2.80$\\
5.00&  8.1   & $-2.11$ & $-2.03$& $-2.47$& $-3.69$& $-2.94$ & $-4.76$&  -&    -&     -       &     -\\
6.00&  7.9   & $-1.86$ & $-1.86$& $-3.70$ & $-6.96$& $-2.84$ & $-5.12$&  -&    -&     -       &     -\\
\hline\hline
\end{tabular}
\label{Table:ejecta}
\end{center}
\end{table*}

\section{Summary and concluding remarks}
\label{sec_discussion}
In this study we investigate formation,  evolution, and mineralogy
of the dust grains that are expected to form in the outflows of TP-AGB stars.
Following the formalism developed by FG06,
a dust-growth model, coupled to a simplified description of the wind dynamics,
is applied to the new TP-AGB evolutionary tracks computed by \citet{marigoetal13}.
The new TP-AGB tracks rely on an accurate treatment of the
molecular chemistry and opacities across the stellar envelope and the
atmosphere,  which  are consistently linked to the changes in the
chemical abundances caused by the third dredge-up and HBB nucleosynthesis.

We pay particular attention to reproduce the measured expansion velocities
and the dust-to-gas ratios of the expanding CSEs around Galactic M and C-stars, and their
observed correlations with the mass loss rates.

As far as the M-stars are concerned, we find
that our computations, based on the original FG06 formulation
for the grain growth, fail the comparison with observations,
predicting terminal velocities that are too low.

The cause lies in the inability of iron-rich silicates
to form and to keep thermally stable in the inner regions of the CSE,
so that the acceleration imparted to the gas
only starts at larger distances and it is insufficient to account for
the observed expansion velocities.

It must be emphasized that the discrepancy between predicted and
measured expansion velocities for  M-stars  is a well-known fact, and it
stems from a more dramatic issue, inherent to the severe inability
of detailed dynamical wind models to even generate an outflow
from O-rich AGB stars, i.e.
to reach dust radiative accelerations larger than the gravitational one
\citep{Woitke06}.

This challenging problem has been more recently tackled by \citet{Hofner08}
and \citet{Bladh12, Bladh13}.
These studies point out that a possible way to produce a wind in
M-stars may rely on photon scattering by large iron-free silicate grains.
Since these grains are transparent to radiation at shorter wavelengths,
they can survive in the inner zones of CSE, where they give
an efficient boost to the gas thanks to their high scattering opacity.

From all these indications, it is clear that in any case, i.e. for any model,
the critical process that should be inhibited is dust destruction in the inner
regions of the CSE.
Guided by this premise, we have addressed the specific question of
predicting higher terminal velocities for the CSEs of M-stars,
so as to eventually match the observations.
Given that in our model mass loss is assumed
and not predicted as in \citet{Hofner08},
our analysis considers the problem from a different
perspective, that is to test the suitability of some
physical assumptions, under the working hypothesis
that the AGB wind does exist.

We have revisited the two mechanisms
that are able to efficiently destroy dust grains, i.e.
in order of importance, chemisputtering by H$_2$ molecules
and sublimation \citep{GS99}.
Concerning chemisputtering, we take into account the suggestion
of several authors that, considering the activation
energy barrier of the reduction reaction of silicates
by H$_2$ (Gardner 1974; Tso \& Pask
1982; Massieon et al. 1993; Tielens, private communication).
the process could be strongly inhibited at the temperatures
and pressures of the regions where dust is predicted to form.
There is also experimental evidence that,
at the pressures typical of a CSE (P$\leq$ 10$^{-2}$dyne~cm$^{-2}$),
chemisputtering may not be efficient \citep{Nagahara96, Tachibana02}.
Furthermore, in recent experiments
condensation temperatures of amorphous silicates
as high as 1350~K have been measured \citep{nagaharaetal09}.

Therefore, we consider it plausible and instructive examining the
case in which the evolution of dust grains is only determined by
growth and free sublimation, hence switching chemisputtering off.
The rate of sublimation in vacuum,
computed following \citet{Kobayashi11},
is used to determine the critical temperature
below which dust can grow, as explained in Section~\ref{sec_subli}.
In this way, we are able to follow the evolution
of the condensation temperature in response to changes in the phyiscal
properties of the CSEs.

The key point is that, without chemisputtering,
the sublimation rates of silicates exceed
their  growth rates  at dust equilibrium temperatures significantly higher than
previously assumed.
In these HCT models the condensation temperature of olivine
is always between 1200~K and 1400~K
and the corresponding gas pressure is always between
P$\sim$10$^{-4}$dyne~cm$^{-2}$ and P$\sim$10$^{-2}$dyne~cm$^{-2}$.
The values of the condensation temperature are in very good agreement with the recent experimental measurements
by \citet{nagaharaetal09} and the value of the pressure corresponds to the condition where
chemisputtering should not be efficient \citep{Nagahara96}.

By neglecting chemisputtering and including a consistent determination
of the condensation temperature,
we find that, even assuming an  opacity typical of dirty silicates,
grains can form and survive in the inner regions of the CSEs (R$\sim$2~R$_*$).
In these regions the grains quickly grow to large sizes, and
provide the gas  with the needed acceleration.
Our HCT models are able to reproduce fairly well the trend
of the velocity with the mass loss rate observed in Galactic M-giants
(Fig.~\ref{v_dotm_M}).

At the lower mass loss rates,
a small mismatch remains that may require further investigation.
A possibility is that those  measured in laboratory evaporation experiments
may not represent the effective sticking coefficients
that regulate the dust growth in CSEs.
For instance, \citet{Nagahara96} noticed that the measured sticking
coefficients refer to the crystalline structure used
in the experiments, and that different structures could
have different energy barriers for the formation reactions,
implying different values of the sticking coefficient.
Other authors have investigated the effects of increasing this coefficient
up to 0.5 \citep{ventura12} for pyroxene, or
even 1.0 for olivine \citep{Hofner_size08}.
We find that just increasing the sticking coefficient of silicates to
$0.2$ is enough to eliminate the residual discrepancy
between observed and predicted  terminal velocities of M-giants.
Another possibility could be that of decreasing the number of initial seeds in
order to reach larger grain sizes that provide a higher opacity.
Lowering the seed number  from 10$^{-13}$ to 10$^{-14}$, the maximum grain sizes
increase from 0.1~$\mu$m to 0.3~$\mu$m, succeeding in reproducing the observed
terminal velocities also for iron-free silicates, as shown in Fig.~\ref{v_dotm_large}.

The predicted dust-to-gas ratios of M-giants are within the observed range.
However though the models cluster around $\delta=1/200$, which is
the value usually adopted for Galactic AGB stars, the observed sample has a median value about
50\% larger, $\delta=1/135$. Furthermore
our models never reach  the highest measured values.
They also follow the usual linear scaling at decreasing metallicity, clustering around $\delta=1/500$
for $Z=0.008$ but again, the few direct data available for LMC M-giants,
indicate that their dust content is more than twice that of our models.

As far as C-star models are concerned, they reproduce fairly well  the observed expansion velocities of Galactic C-stars.
In the homogeneous growth scheme, at the base of our carbon dust formation model,
the key role is played by the carbon over oxygen excess  $\epsilon_{\rm C}-\epsilon_{\rm O}$
\citep{Mattsson10}. This parameter is mainly determined by the complex interplay
between mass loss, third~dredge-up and HBB.
For example, we have shown that delaying the super-wind phase
without suitably lowering  the efficiency of the third~dredge-up,
the carbon enrichment at the surface would be larger,
and consequently the terminal velocities would be higher.
We have also shown that varying the sticking coefficient of carbon
can induce some degeneracy with the metallicity but, unless one assumes
a very low value ($\sim$0.1), the velocities of Galactic C-stars can be well reproduced.
We compare the dust-to-gas ratios of the C-star models with three different Galactic samples.
Contrary to those of M-giant models which show little dependence on the mass loss rates,
those of C-stars increase at increasing mass loss rates.
This trend is present only in one of the three samples \citet{Groenewegen98}
while it is absent in the other two. At low mass loss rates the median values of the three samples
are similar and they are reproduced by our models.
At the high mass loss rates the models are not able to reproduce the largest observed values.
We notice also that the maximum mass loss rate obtained by our models is about 2-3 times lower than the
maximum observed one.

The coupling of the homogeneous growth scheme to
our AGB models gives rise to the following trends with decreasing metallicity.
The velocities at a given mass loss rate increase, the dependence of the dust-to-gas ratios
on the mass loss rate becomes stronger but their values at low mass loss rates decrease.
Unfortunately, in spite of the huge literature on C-stars at metallicities lower than solar,
there is a severe lack of {\it direct} measurements of dust-to-gas ratio and terminal velocities.
The velocities of six C-stars toward the Galactic Halo \citep{Lagadecetal12} are clearly
significantly lower than those of
the bulk of our models at $Z=0.001$ and even at $Z=0.008$. However, for the three objects  likely belonging to the thick disk,
we can reproduce dust-to-gas ratios and velocities with C-stars of $Z=0.008$ and M=1.5~M$_\odot$.
For the other three stars, that were classified as Halo members, the data can be reconciled with
models with  $Z=0.001$ and M$\approx$4~M$_\odot$, where the carbon excess is lowered by
an efficient HBB. This value for the mass is somewhat higher than that estimated by
 \citep{Lagadecetal12}, on the basis of their luminosity (M$\approx$2-3~M$_\odot$).
Furthermore, and perhaps more important, the particularly strong equivalent width of
their 7.5~$\mu$m C$_2$H$_2$ feature
is difficult to reconcile with a relatively lower carbon excess.
This could be an indication that the efficiency of carbon dust formation decreases at decreasing metallicity
as suggested by \citet{vanloon08}. Whether the efficiency becomes
lower by assuming an heterogeneous nucleation on metallic seeds (TiC, ZrC and MoC) needs still to be investigated.

For C-stars we have also compared the predicted
abundance ratios of SiC relative to carbon,
with the observations.
At solar metallicity our LCT models predict values that are significantly higher
than those observed among Galactic C-stars while at intermediate metallicity
the agreement is fairly good. The three Halo stars at low metallicity have measured SiC/C ratios
that are too high with respect to our models. But this is more likely a intriguing
problem related to their origin because the mid infrared SiC feature is observed to decrease significantly
at decreasing metallicity. In any case the SiC/C ratio may depend
on effects other than the global metallicity or the
carbon excess. For example, we have shown that increasing the condensation temperature of carbon,
lowers significantly the SiC/C ratios of models with moderate and low mass loss rates.

Our simple models provide a powerful check
of the internal processes that regulate the evolution of  TP-AGB stars,
that should be added to all other existing observable tests
used to calibrate this important phase of stellar evolution.

In this work we also present new  dust ejecta of TP-AGB stars,
computed for three initial metallicities,
$Z=0.001,\, 0.008$, and $0.02$, and a few values of the initial mass,
from 1~M$_{\odot}$ to $\simeq 5-6$~M$_{\odot}$.

At low metallicity,  the bulk of the dust consists of amorphous carbon, whereas,
at increasing metallicity, the range of initial stellar masses producing
this dust species  is limited between 2~M$_\odot$ and 4~M$_\odot$ at $Z=0.02$.
Silicate dust production dominates
at lower masses due to weak or absent dredge-up and, at higher masses,
because of efficient HBB.
The ejecta depend weakly  on the details
of the adopted condensation prescriptions.
This is true not only for silicate dust, but also for carbon dust as
demonstrated with the aid of test calculations
in which we  artificially increased the
gas temperature threshold for the activation of the C$_2$H$_2$ chains.

Differently, the results do depend on the underlying TP-AGB evolutionary models.
In particular,  a sizeable discrepancy affects the predictions for the carbon dust production
at low metallicity,  between \citet{FG06} and our new results
based on the \texttt{COLIBRI} code on one side,
and the results of \citet{ventura12} on the other side.
Furthermore, silicate ejecta computed with the  \texttt{COLIBRI} code
are significantly larger than those of \citet{FG06}  and \citet{ventura12}.
For C-stars these differences  are related to our poor understanding of the
mass loss and third~dredge-up processes,
since their ill-determined efficiency affects dramatically the surface C/O ratio.
Differences in silicate ejecta are more subtly related also to
initial metal partitions and adopted opacities, so that it is difficult to trace back
their origin.

In both full evolutionary TP-AGB  models \citep{ventura12},
and TP-AGB models based on numerical envelope integrations  \citep{marigoetal13}
mass loss and third dredge-up
are described by means of suitable parameters.
For mass loss, both models use empirical laws taken from the literature,  while
the third~dredge-up  is regulated by a parametrized efficiency
of convective overshoot in full models, or directly described
by its efficiency parameter in envelope-based models.
How these parameters depend on the ambient metallicity is one of the
far-reaching questions, that are crucial for the interpretation of
extragalactic observations up to high redshift.

Indeed, the major aim of the  \texttt{COLIBRI} code, together with
the piece of work presented here, is
to provide a fast and flexible tool able to put different observations
inside a common interpretative framework.
In this respect,  the dust formation scheme described here
will be soon applied to study the emission properties
of the circumstellar dusty envelopes \citep{Bressan98, Marigo_etal08}.

\subsection*{Acknowledgements}
We thank the referee, Jacco van Loon, for his careful reading of the manuscript and his many
suggestions that helped us to improve the paper.
We also thank L. Danese, A. Tielens, H. Kobayashi, I. Cherchneff and L. Agostini for the fruitful discussions.
We acknowledge financial support from contract ASI-INAF I/009/10/0,
and from the Progetto di Ateneo 2012,
CPDA125588/12 funded by the University of Padova.
AB acknowledges financial support from MIUR 2009.

\bibliographystyle{mn2e/mn2e}
\bibliography{dust}
\label{lastpage}
\end{document}

%% file: journalDefs.tex
\def\apjl{ApJ}%
\def\apjs{ApJS}%
\def\ao{Appl.~Opt.}%
\def\aap{A\&A}%